**Broadband gate-tunable THz plasmons in graphene heterostructures**

Baicheng Yao[1,2,3*], Yuan Liu[4,5], Shu-Wei Huang[1], Chanyeol Choi[1], Zhenda Xie[1], Jaime Flor Flores[1], Yu Wu[2], Mingbin Yu[6], Dim-Lee Kwong[6], Yu Huang[4,5], Yunjiang Rao[2], Xiangfeng Duan[5,7*], and Chee Wei Wong[1*]

[1]Fang Lu Mesoscopic Optics and Quantum Electronics Laboratory, University of California, Los Angeles, CA 90095, United States
[2]Key Laboratory of Optical Fiber Sensing and Communications (Education Ministry of China), University of Electronic Science and Technology of China, Chengdu 610054, China
[3]Cambridge Graphene Center, University of Cambridge, CB3 0FA, United Kingdom
[4]Department of Materials Science and Engineering, University of California, Los Angeles, CA 90095, United States
[5]California Nanosystems Institute, University of California, Los Angeles, CA 90095, United States
[6]Institute of Microelectronics, Singapore 117685, Singapore
[7]Department of Chemistry and Biochemistry, University of California, Los Angeles, CA 90095, United States
* Correspondence to: by252@cam.ac.uk; xduan@chem.ucla.edu; cheewei.wong@ucla.edu

**Graphene, a unique two-dimensional material of carbon in a honeycomb lattice [1], has brought remarkable breakthroughs across the domains of electronics, mechanics, and thermal transport, driven by the quasiparticle Dirac fermions obeying a linear dispersion [2-3]. Here we demonstrate a counter-pumped all-optical difference frequency process to coherently generate and control THz plasmons in atomic layer graphene with an octave tunability and high efficiency. We leverage the inherent surface asymmetry of graphene for a strong second-order nonlinear polarizability $\chi^{(2)}$ [4-5], which together with tight plasmon field confinement, enables a robust difference frequency signal at THz frequencies. The counter-pumped resonant process on graphene uniquely achieves both energy and momentum conservation. Consequently we demonstrate a dual-layer graphene heterostructure that achieves the charge- and gate-tunability of the THz plasmons over an octave, from 9.4 THz to 4.7 THz, bounded only by the pump amplifier optical bandwidth. Theoretical modeling supports our single-volt-level gate tuning and optical-bandwidth-bounded 4.7 THz phase-matching measurements, through the random phase approximation with phonon coupling, saturable absorption, and below the Landau damping, to predict and understand the graphene carrier plasmon physics.**

The discovery of graphene spurred dramatic advances ranging from condensed matter physics, materials science to physical electronics, mechanics, and thermal processes. In optics [6-7], the additional chiral symmetry of the Dirac fermion quasiparticles of graphene [8] enables an optical conductivity defined only by the fine structure constant $\pi\alpha$ [9], one that is remarkably charge-density tunable [10-11] and with broadband nonlinearities [12-15]. The collective oscillations of the two-dimensional correlated quasiparticles in graphene [16] naturally make for a fascinating cross-disciplinary field in graphene plasmonics [17], with applications ranging from tight-field-enhanced modulators, detectors, lasers, polarizers, to biochemical sensors [18-22]. Different from conventional noble metal plasmons, graphene plasmons are dominant in the terahertz and far-infrared frequencies [23]. To excite and detect these plasmons, specialized techniques such as resonant scattering nanoscale antennae near-field microscopy or micro- and nano-scale scattering arrays have been pursued, albeit still using terahertz/far-infrared sources [24-28]. Recently nonlinear optical processes, only with free-space experiments, have proven especially effective in generating graphene plasmons with efficiencies up to $10^{-5}$ [4-5]. However, to date, it is challenging to generate, detect, and control on-chip graphene plasmons all-optically, a key step towards planar integration and next-generation high-density optoelectronics.

Concurrently THz generation has recently been revisited by a number of studies for imaging, spectroscopy, and communications [29]. While a wide tunability in THz can provide new grounds for broadband stand-off spectroscopy and wavelength-agile ultrahigh-bandwidth communications, tunability in THz materials has so far been limited (Supplementary Table S1). Here we demonstrate experimentally the charge- and gate-tunability of THz plasmons over a full octave, from 9.4 THz to 4.7 THz, bounded only by the pump amplifier optical bandwidth. Through the surface asymmetry of dual-layer graphene heterostructures and tight plasmon field confinement, we leverage the intrinsically strong $\chi^{(2)}$ for difference-frequency coherent THz generation. We implement a chip-integrated counter-pumped resonant process for frequency- and phase-matching over a full octave of the full-scale THz carrier frequency. Our designed heterostructure achieves the widely-tunable THz generation via gating at the single volt level, matching our conductivity models and numerical predictions.

Figure 1a shows the graphene on silicon nitride waveguide (GSiNW) architecture examined in this work. The GSiNW has a bottom atomic layer graphene connecting the drain and source contacts, a layer of alumina working as a thin dielectric barrier, and a second atomic layer of top graphene connecting the gate. The silicon nitride waveguide has a 1 μm width × 725 nm height, and the graphene-$Al_2O_3$-graphene hybrid heterostructure is assembled with direct contact to the nitride core, enabling effective light-graphene interaction along the ≈ 80 μm waveguide overlap region. The waveguide input-output regions are tapered for effective on/off-chip coupling. Detailed nanofabrication is shown in Supplementary Section S3. The graphene layers serve simultaneously as the active electrodes, the second-order nonlinearity media and the nanoscale plasmon waveguides. Surface asymmetry of graphene has an effective second-order nonlinear polarizability $\chi_{eff}^{(2)}$ described approximately as

$$\chi_{eff}^{(2)} = \frac{e^3}{4\pi\hbar^2} \frac{1}{k_{SP}\sqrt{f_s f_p}} \left[ \frac{\pi}{2} + \arctan\left( \frac{2\pi\sqrt{f_s f_p} - 2v_F k_F}{\gamma} \right) \right] \qquad (1)$$

where $f_s$, $f_p$ are the frequencies of the pump and the signal laser, $e = -1.6 \times 10^{-19} C$ is the unit charge, $\hbar$ is the reduced Plank's constant, $\gamma$ is the scattering rate of graphene, $v_F$ is the Fermi velocity, $k_F = (2m_e E_F)^{1/2}$ is the Fermi momentum, and $k_{SP}$ is the counter-pumped phase-matched momentum. The detailed derivation is shown in Supplementary Section 2. We launch the pump and signal lasers into the GSiNW in opposite directions as shown in Figure 1.

The on-chip difference frequency generation (DFG) process for the THz plasmon generation is shown in both Figure 1a inset and Figure 2a. Driven by the $\chi_{eff}^{(2)}$ polarizability, the energy of one pump photon is distributed into the lower-energy signal photon and a plasmon. In the GSiNW, the photonic modes propagate along the $Si_3N_4$ core in opposite directions while the plasmonic mode co-propagates along the graphene interface in the same direction as the pump. Our counter-pumped nonlinear phase matching scheme [30] satisfies both energy conservation $f_{SP} + f_s = f_p$ and momentum conservation $k_{SP} = k_p + k_s$, where $f_{SP}$, $f_s$, and $f_p$ are the plasmon, signal and pump frequencies respectively. $k_{SP}$, $k_s$, and $k_p$ are the momenta of the plasmon, signal and the pump respectively. Based on the optical dispersion $k = 2\pi f n/c$, phase matching condition is achieved when $f_s n_s - f_{SP} n_{SP} = -f_p n_p$, where $n_p$, $n_s$, and $n_{SP}$ are the effective indexes of the plasmon, signal, and pump respectively. In addition, we note that both the optical pump and signal wavelength modes propagating along the graphene-$Al_2O_3$-graphene-$Si_3N_4$ heterostructure are transverse magnetic (TM) polarized. Optical TM polarization enables strong evanescent field interactions with the graphene layers and both being TM enables the DFG plasmon generation.

The surface plasmon polariton frequency $f_{SP}$ is determined by the Fermi level based graphene dispersion, with a plasmon frequency each for the top and bottom graphene layers. First consider the case of zero gate voltage. In our dual-layer graphene-$Al_2O_3$-graphene capacitor at zero gate voltage, both the top and bottom layer graphene are intrinsically positively charged (p-doped; top Fermi level $E_{T0} \approx$ bottom Fermi level $E_{B0} \approx$ - 50 meV) due to carrier trapping. With the small (30 nm) interlayer distance between the top and bottom layer graphene, the plasmon modes weakly couple to form two hybrid modes, symmetric and antisymmetric. In the low frequency regime, the dispersions of the symmetric and antisymmetric modes are consequently described as:

$$f_{sym} = \frac{1}{2\pi}\sqrt{\frac{2e^2}{\varepsilon}(E_T + E_B)k_{SP}} \qquad (2)$$

$$f_{asym} = \frac{1}{2\pi}\sqrt{\frac{4e^2 E_T E_B d}{\varepsilon(E_T + E_B)}k_{SP}} \qquad (3)$$

Here $\varepsilon$ is the background permittivity and $d$ is the dielectric layer thickness. The symmetric mode is observed in our measurements while the anti-symmetric mode is strongly damped and hence hard to observe experimentally. Detailed numerical calculations are shown in Supplementary Section S2.

Next we consider the case of applied gate voltages ($V_G$). The Fermi levels of the top and bottom graphene are significantly and oppositely tuned, with $|E_T| = |E_{T0} + E_G|$ and $|E_B| = |E_{B0} - E_G|$ where $E_G$ [$= \hbar|v_F|(\pi N)^{-1/2}$] is the quasi-Fermi level determined by the gate-injected electron density $N$ and Fermi velocity $v_F$. This modulates the plasmon mode dispersion – and hence DFG phase matching – with an order-of-magnitude larger modulation than the mode-splitting frequency change at zero voltage. The difference in the top and bottom graphene Fermi levels (when gated) leads to different top and bottom plasmon frequencies, which thus has negligible interlayer plasmon coupling. Figure 1b shows the top-view optical micrograph of the nanofabricated source-drain and gate electrodes on the graphene-$Al_2O_3$-graphene hybrid heterostructure. The top and bottom graphene layers are denoted by white and blue dashed boxes respectively, the selected core waveguide denoted by the red dashed line, and the gold contact electrodes illustrated (with the top gate brighter than the source and drain electrodes as it is patterned above the $Al_2O_3$ dielectric).

The DFG experimental setup consists of a mode-locked picosecond pump pulse at 39.1 MHz repetition rate launched into the GSiNW from the left, with an amplified continuous-wave (cw) signal counter-launched from the right, both in TM polarizations. The detailed experimental setup is shown in Supplementary Section S4. Figure 1c shows an example time-domain modulated transmission, with ≈ 200 W pump peak power and at 1532 nm. With graphene saturable absorption, the broadband transmission of the high-peak-power pulsed pump is ≈ 2.1 dB higher than under a cw pump of the same average power. With the modulated pump pulses together with a counter-pumped cw signal, the resulting observed transmission envelope waveform is sinusoidal with the saturable absorption induced modulation. The modeled saturable absorption transmission and modulation are also shown in the black solid curves of Figure 1c.

As noted in Figure 1a, we proceed to search for the plasmon generation experimentally by monitoring the signal intensity as a function of the swept cw signal frequency, with the pulsed pump laser switched on. With the presence of the plasmon in the DFG phase matching process, the signal intensity will rise as shown in Figure 1d left panel ($\Delta I_{DFG}$). In this case, the pump laser is fixed at 1532 nm (195.82 THz) and the cw signal laser is swept from 1570 nm to 1610 nm (191.08 THz to 186.34 THz) at 1 nm/ms scanning rate. To directly detect this DFG plasmon signal over noise, we implemented a 100 kHz modulation on the mode-locked picosecond laser, with lock-in filtering, amplification and balanced detection. In Figure 1d left panel, the measurements were done at $V_G = 0$ V. The plasmon is detected when the signal photon is at 1593.2 nm (188.4 THz) in this case – this corresponds to a plasmon frequency $f_{SP}$ of 7.5 THz. The pump on-off intensity contrast ratio is measured to be $\approx 1.7$ a.u. (with the pump off lock-in signal referenced to zero), which arises from the residual saturable absorption modulation as mentioned above [21]. With the presence of the DFG plasmon, an additional 0.3 a.u. peak intensity contrast ($\Delta I_{DFG}$) is observed. In Figure 1d right panel, when $V_G = -0.7$ V, we observe two enhanced peaks at 1593.7 nm (188.2 THz) and 1607.2 nm (186.7 THz). This corresponds to a top and bottom graphene plasmon frequency at $\approx 7.6$ THz and 9.2 THz respectively.

With the gate voltage applied, the Fermi level is tuned from $E_F$ to $E_F'$ and $f_{SP}$ changes to be $f_{SP}'$. Correspondingly, the enhanced signal $f_s$ shifts to $f_s' = f_p - f_{SP}'$. We measure the correlation of $E_F$ with $V_G$ through the $I_{SD}$-$V_G$ measurements, as shown in Figure 2b. Here $I_{SD}$ is the source-drain current and $R_{SD}$ is the source-drain resistance. In our chip, when tuning $V_G$ up to $\pm 4$ V, $I_{SD}$ changes from 1.29 µA to 1.47 µA; correspondingly, $R_{SD}$ is in range of 6.9 kΩ to 7.8 kΩ. (At $V_G = 0$, $E_{T0} = E_{B0} = -50$ meV due to the natural doping) When close to the Dirac point ($V_{Dirac} = 0.25$ V), graphene has the highest sheet resistance. Accordingly, $|E_F|$ is estimated to be tuned in range of 0 to $\approx 270$ meV. When $V_G$ approaches 0.25 V (-0.25 V), $E_F$ of the bottom (top) graphene reaches close to the Dirac point (Supplementary Section S4.4). Furthermore we note that, when $E_F$ changes, the effective second-order nonlinear susceptibility $\chi_{eff}^{(2)}$ of the graphene also changes, as shown in Eq. (1). Consequently, the plasmon intensity ($I_{SP}$) and signal intensity enhancement ($\Delta I_{DFG}$) would be gate-tuned as well.

With the gate tunability of $E_F$, we observe the tuning of the graphene THz plasmon signal, as shown in Figure 2c. In the top panel of Figure 2c, when $V_G$ increases from -0.7 V to -0.3 V, the DFG enhanced signal peak of the bottom layer ($\lambda_{s,Bot}$) blue-shifts from 1607.2 nm to 1601.3 nm, with intensity increasing from 0.28 to 0.37 a.u (at $V_G = -0.4$ V), then decreasing back to $\approx 0.32$ (at $V_G = -0.3$ V). Simultaneously $\lambda_{s,Top}$ and intensity of the top layer decreases significantly. In the bottom panel, as we further increase $V_G$ from -0.2 V to 0.2 V, the signal peak generated by the bottom layer graphene plasmons blue-shifts from 1598.6 nm to 1578.1 nm, with intensity decreasing from 0.35 to 0.09 a.u. However, the signal peak generated by the top layer graphene plasmons begins to reappear from -0.05 V and then red-shifts to 1601.6 nm with intensity increasing to 0.11 a.u. In Figure 2c, the measured $\Delta I_{DFG}$ of bottom (top) layer graphene is 0.37 a.u. at $V_G \approx -0.4$ V (0.11 a.u. at $V_G \approx 0.4$ V). The $\Delta I_{DFG}$ of the top layer graphene is $\approx 3$ times weaker than the $\Delta I_{DFG}$ of the bottom layer, because the top layer graphene is further from the waveguide core, where the evanescent field overlapping of the pump and signal are weaker.

Figure 2d analyzes the observed gate tuning of the THz graphene plasmons. Here the blue dots and red diamonds are the measured results while the solid curves are the theoretical fittings. With the bottom and top layer graphene having the same carrier densities (e.g. $n_{electron}$ on the bottom layer equals $n_{hole}$ on the top layer), they have symmetrical curves. Limited by the *L*-band amplifier spectral window and the ±0.04 a.u. noise, plasmons with $f_{SP}$ higher than 9.4 THz or lower than 4.7 THz are difficult to determine rigorously. We note that our demonstrated plasmon tuning range from 4.7 THz to 9.4 THz already spans over an octave. Based on the measured results, Figure 2d also provides the tunable effective index $n_{SP}$ and the tunable wavelength $\lambda_{SP}$ of the plasmons, by referring the dispersion relationship $n_{SP} = (f_p n_p + f_s n_s)/(f_p - f_s)$ and $\lambda_{SP} = c/(f_{SP} n_{SP}) = c/(f_p n_p + f_s n_s)$. Here $f_p$ is fixed at 195.8 THz. For the bottom layer graphene when $V_G$ approaches $V_{Dirac}$, $n_{SP}$ increases from ≈ 69 to 116, correspondingly, $\lambda_{SP}$ changes in range of 460 nm to 466 nm during gate voltage modulation. The top layer graphene has a symmetrical $n_{SP}$ - $V_G$ measured dependence. This result supports that a lower Fermi level (closer to the Dirac point) could bring a better plasmonic confinement.

To further understand the gate tunable plasmon generation, we investigate its phase matching conditions in Figure 3. To generate the THz graphene plasmons, phase matching conditions of the counter-pumped DFG and dispersion of the plasmonic modes must be satisfied simultaneously. By using random phase approximation (RPA), we map the graphene plasmon dispersion at $V_G$ = 0 V in Figure 3a and 3b, here both the top layer and the bottom layer graphene have the same $|E_F|$. The photon-electron interaction loss $L_{SP}$ is normalized. In the regime with the graphene $f_{SP}$ much lower than the Landau damping, the dispersion of graphene plasmons $k_{SP}(f_{SP})$ behaves approximately as a quadratic function ($k_{SP} \propto f_{SP}^2$). In the $f_{SP}$-$k_{SP}$ band structure map, the phase matching condition of the DFG could be written as $(c/2\pi)k_{SP} = -f_{SP} n_s + f_p(n_p + n_s)$, shown as the grey solid line plots (near-vertical lines) of Figure 3a and 3b. Here $n_s$ and $n_p$ are the effective indexes of the pump and signal. Hence, in Figure 3a and 3b, plasmons are generated only at the intersections of the graphene dispersion curves and the DFG phase matching lines.

At $V_G$ = 0 V, the Fermi levels of the top and bottom graphene layers are nearly the same. As a result, $f_{SP}$ of both the top and the bottom graphene layers are essentially the same, at ≈ 7.5 THz in experiment, as the measured blue dots show in Figure 3b. This is ≈ 0.3 THz higher than the numerical calculation (green dashed curve) that did not consider the weak interlayer plasmon coupling between the bottom and top graphene layers. When this interlayer plasmon coupling is taken into account in the numerical calculations, the green dashed curve moves up to the blue solid line in Figure 3a and 3b. This is the symmetric mode of the dual-layer graphene (the antisymmetric mode has strong damping and hence hard to observe). There is good match between the measured blue dots and the numerically calculated blue solid line. Detailed theoretical discussions are provided in Supplementary Section S2.4. As a comparison and verification, we also measure $f_{SP}$ at $V_G$ = 0 V in another GSiNW sample, with a 60 nm $Al_2O_3$ layer which then has negligible interlayer coupling. This is shown in the green triangles, matching well with the green dashed curve. Details of the 60 nm $Al_2O_3$ dielectric device are also shown in Supplementary Figure S5.3.

Moreover, to verify $f_{SP}$ matches the plasmonic dispersion curve, we change the $f_p$ to vary the DFG phase matching points, as illustrated in the zoomed-in Figure 3b. By tuning $\lambda_p$ from 1532 nm to 1542 nm (195.8 THz to 194.6 THz), we observe the enhanced signal peak location $\lambda_s$ changes from 1593.2 nm to 1603.0 nm (188.3 THz to 187.2 THz). Hence, $f_{SP}$ decreases from 7.5 THz to 7.4 THz: the trace follows the graphene plasmonic dispersion well. The measured spectra and $\lambda_p$ - $\lambda_s$ correlation is described in Supplementary Section 5.4. When we tune the gate voltage, the dispersion curves of the bottom and top layer graphene move independently. As a result, the generated plasmons on the bottom and top atomic layers have different $f_{SP}$, supporting the results of Figure 2. For instance, Figure 3c and 3d shows the scenarios of $V_G$ = -0.7 V and $V_G$ = 0.2 V. Here $f_p$ is fixed at 195.8 THz, blue dots and red diamonds show the measured results of the bottom and top layer graphene respectively (notice the change in polarity of $V_G$ swaps the bottom and top $f_{SP}$s).

We next examine the intensity of the DFG plasmons ($I_{SP}$). In the DFG process, $I_{SP}$ is proportional to the intensity of the pulsed pump $I_p$ and the intensity of the cw signal $I_s$, as:

$$I_{SP} = \frac{(\chi_{eff}^{(2)})^2 I_p I_s}{L_{SP}^2} \tag{4}$$

Detailed theoretical derivations are shown in Supplementary Section S2.3. In the experiment, $I_{SP}$ could be directly estimated from the measured $\Delta I_{DFG}$. Applying the Manley-Rowe relation, *i.e.* the conservation of photon numbers, we can rewrite this relationship as:

$$I_{SP} = \Delta I_{DFG} \frac{f_{SP}}{f_p - f_{SP}} \tag{5}$$

By fixing the signal intensity at 1.4 W and $f_p$ at 195.8 THz, we illustrate the spectra of the $\Delta I_{DFG}$ in Figure 4a and 4b. In Figure 4a, $V_G$ = 0 V, while in Figure 4b, $V_G$ = -0.7 V. For either the bottom or top layer graphene, $\Delta I_{DFG}$ increases linearly when $I_p$ is increased from 0 to 32 mW. The insets of Figure 4a and 4b summarize the $\Delta I_{DFG}/\Delta I_p$ correlations, with a slope on the order of $10^{-3}$ a.u./mW level. Considering the optical loss and amplifications, we estimate the plasmons generated on-chip are on the order of single nWs.

We further examine the conversion efficiency of the second-order nonlinearity based plasmon generation in Figure 4c and 4d. Here we define the conversion efficiency $\eta = I_{SP}/I_s I_p$. From Eq. (1), by normalizing pump power, $\eta = (\chi_{eff}^{(2)}/L_{SP})^2$. For $f_p \approx f_s \gg f_{SP}$, $\eta$ could be approximately written as

$$\eta = \frac{e^3}{2\pi^3 \hbar^2 k_{SP} \sqrt{f_s f_p}} \left\{ \frac{\pi}{2} + \arctan\left(\frac{2\pi\sqrt{f_s f_p} - 2v_F\sqrt{2m_e E_F}}{\gamma}\right) \right\} \arctan\left(\frac{E_F}{\varepsilon\gamma}\right) \tag{6}$$

Here $m_e$ is the electron rest mass, $\gamma$ is the scattering rate, $\hbar$ is reduced Plank's constant, and $v_F$ is the Fermi velocity at $\approx 10^6$ m/s. In the GSiNW, $\chi_{eff}^{(2)}$ and $1/L_{SP}$ are of opposite trend: when $E_F$ = 0 eV, graphene has the largest $\chi_{eff}^{(2)}$, however its carrier density is minimal and $L_{SP} \rightarrow \infty$: graphene plasmon is completely damped. When $E_F$ is high, due to the high carrier density, $L_{SP}$ could be low but $\chi_{eff}^{(2)}$ approaches 0. As a balanced trade-off, with increasing Fermi level, $\eta$ rises first, and then drops gradually when $E_F > 130$ meV. That means, by selecting a proper $|V_G - V_{Dirac}| \approx 0.5$ V to ensure $E_F \approx 130$ meV we can find a highest $\eta \approx 6\times10^{-5}$ W$^{-1}$. Under a tuned $E_F$, $\eta$ of the bottom and top graphene layers can indeed be degenerate.

In this work, by using counter-pumped $\chi^{(2)}$ difference frequency-generation, we demonstrate THz plasmon generation and control in chip-scale integrated graphene for the first time. With the dual-layer graphene heterostructure, our counter-pumped configuration enables phase and frequency matching, with robust DFG signal detection through lock-in and balanced detection. The coherent DFG excitation is gate-tunable for both the graphene layers, with symmetric-antisymmetric frequency crossing between the two layers and with tunability from 4.7 to 9.4 THz, a nearly 50% full-scale tunability. Under positive and negative gate voltages, we observe the plasmon dispersion frequencies exchange between the top and bottom graphene layers, with a linewidth quality factor that increases slightly with $f_{SP}$, up to $\approx 60$. The optimal planar THz plasmon generation efficiency approaches $10^{-4}$ when the Fermi level is 0.13 eV, drawing from a trade-off with larger $\chi^{(2)}$ but larger plasmon loss from phonon coupling ($L_{SP}$) with smaller $E_F$. These observations on the chip-scale graphene THz plasmon generation and control open a new architectural platform for widely-tunable THz sources, gate-tunable metasurfaces, and two-dimensional atomic crystal optoelectronics.

**Methods**

**Plasmon generation via counter-pumped surface $\chi^{(2)}$ nonlinearity.** Detailed theoretical analysis is provided in Supplementary Section S2. It describes the dispersion of the silicon nitride waveguides, phase matching conditions on the GSiNWs, graphene index modulation, the DFG process and nonlinear conversion efficiency, and the dual-layer graphene plasmonic coupling.

**Nanofabrication of the graphene based semiconductor chip.** The top oxide cladding of the silicon nitride waveguide core is chemically etched with wet buffered oxide etch to increase the evanescent field coupling to graphene. After etching, the distance between the core to the top surface is less than 20 nm, ensuring good light-graphene interaction. Then, a monolayer graphene grown by chemical vapor deposition (CVD) is transferred onto the chip using conventional wet transfer technique, followed by patterning by photolithography and oxygen plasma etching. This graphene layer is regarded as the bottom layer graphene with a size of 100 μm × 40 μm. After graphene transfer, a Ti/Au (20/50 nm) pad is deposited using e-beam evaporation, serving as source-drain electrodes. By using the source and drain, the resistance of the bottom layer graphene could be measured. Subsequently, a thin 40 nm layer of Al$_2$O$_3$ is deposited using atomic layer deposition, providing sufficient capacitance for the graphene based semiconductor chip. On top of the $Al_2O_3$ insulator, another graphene layer is covered, linked to the gate electrode. The fabrication process of the graphene based semiconductor chip is shown in Supplementary Figure S3.1. Characterization of the nanofabrication is shown in Supplementary Section S3.2.

**Experimental arrangement.** To enable detection of DFG plasmon signal, four experimental techniques are implemented: **(i)** Both the pump and the signal are TM polarized, maximizing the graphene-light interaction; **(ii)** A mode-locked picosecond pulsed laser serves as the pump, allowing the maximum effective energy density in the GSiNW to reach 50 mJ/cm$^{-2}$, ensuring that the DFG process is fully excited; **(iii)** The cw signal is amplified to have a maximum power of 1.6 W, which can effectively pre-saturate the graphene layers, further increasing the surface $\chi^{(2)}$ efficiency, decreasing the loss, and reducing the effect of a high-peak-power pulse-induced optical modulation; **(iv)** Balanced photodetection (BPD; New Focus 2017) and lock-in amplification (Stanford Research Systems 830) are implemented so that the original DC background of the signal is suppressed, with the balanced signal further filtered and integrated (up to 100 μs) in a lock-in amplifier, amplifying the selected signal dramatically (up to 60 dB) and suppressing white noise effectively. The bandwidth of our BPD and lock-in amplifier is ≈ hundreds kHz, and hence we use a 100 kHz sinusoidal waveform to modulate the pump pulses in a slow envelope. Details of the experimental architecture are shown in Supplementary Section S4.

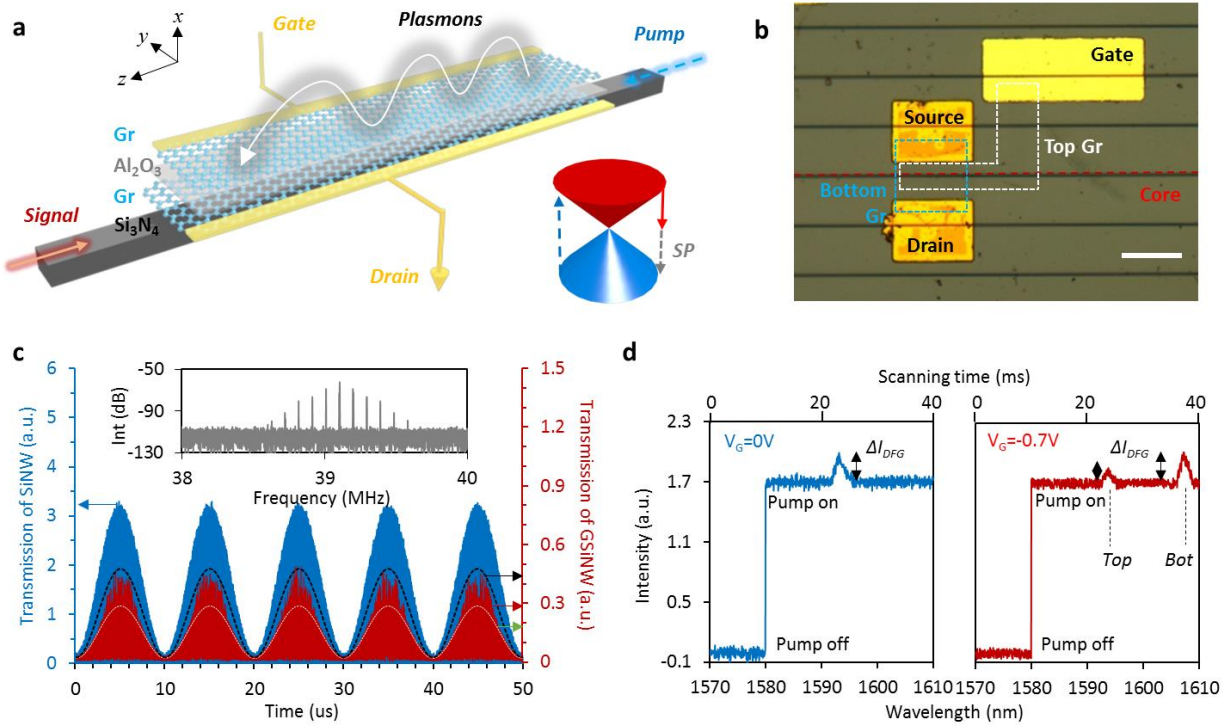

**Figure 1 | Generating and controlling THz plasmons in graphene heterostructures via counter-pumped all-optical nonlinear processes. a,** Schematic illustration of the dual-layer graphene hybrid for difference frequency generation (DFG) via counter-pumped surface $\chi^{(2)}$ nonlinearity. Dual-layer graphene is deposited onto a silicon nitride waveguide core (GSiNW) with an $Al_2O_3$ spacer. Inset: Dirac cone structure of the DFG process. **b,** Top-view optical microscope image of the GSiNW. Bottom and top graphene layers are marked by the blue and white dashed boxes respectively. Dark left-right horizontal lines are the $Si_3N_4$ waveguides (the selected one is marked by red dashed line), orange rectangles are the sources and drain contacts, and the gate contact on the surface is bright yellow. Scale bar: 50 μm. **c,** Measured optical transmissions: blue curve shows the SiNW without the graphene layers; red curve shows the GSiNW; white dashed curve and black dashed curve illustrates the modeled linear transmissions. Inset: modulated pump RF spectrum, with the 39.1 MHz peak from the mode-locked pulse train, while the 100 kHz sideband harmonics are from the sinusoidal modulation. **d,** The measured DFG-based signal enhancement in the optical spectra. With $f_s$ scanning, DFG induces an additional 0.3 a.u. measured enhancement, as marked by $\Delta I_{DFG}$. When $V_G$ = 0 V (left panel), a distinct peak is observed at 1593.2 nm. When $V_G$ = -0.7 V (right panel), two enhanced DFG peaks are observed separately (1593.7 nm and 1607.2 nm), corresponding to the top and bottom layer graphene plasmons respectively.

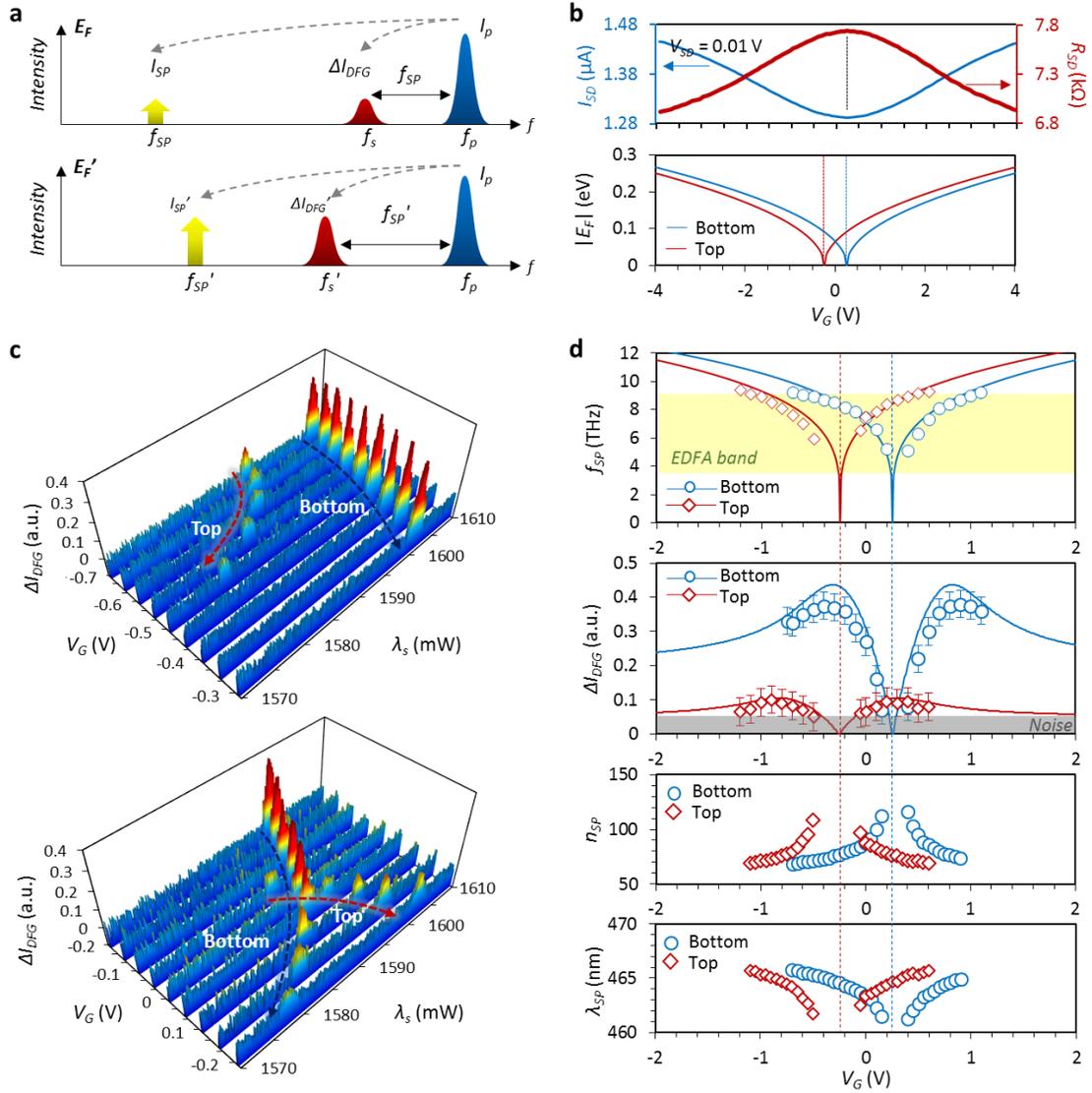

**Figure 2 | Observations and gate tunability of the DFG plasmons generated on graphene. a,** Nonlinear process of the DFG. **b,** Top panel shows the measured $I_{SD}$-$V_G$ correlation and sheet resistance of the GSiNW (blue dots), under a fixed $V_{SD}$ of 10 mV. Bottom panel shows the derived Fermi level of the bottom layer graphene (blue) and the top layer graphene (red), based on the measured $I_{SD}$-$V_G$ correlation. **c,** Measured spectra of the $\Delta I_{DFG}$. Here the $V_G$ mapping step is 50 mV. **d,** Gate tunable parameters of the DFG plasmons, from top to bottom: the observed frequency $f_{SP}$, the intensity $\Delta I_{DFG}$, the effective index $n_{SP}$, and the wavelength $\lambda_{SP}$, respectively, as well as a function of gate voltage $V_G$. In panels of **d,** the solid curves are the theoretical fittings, the band limitation of our *L*-band EDFA (1570 nm to 1610 nm) is marked by the yellow region, and the background noise up to ~ ±0.04 a.u. is marked by a grey area and the error bars.

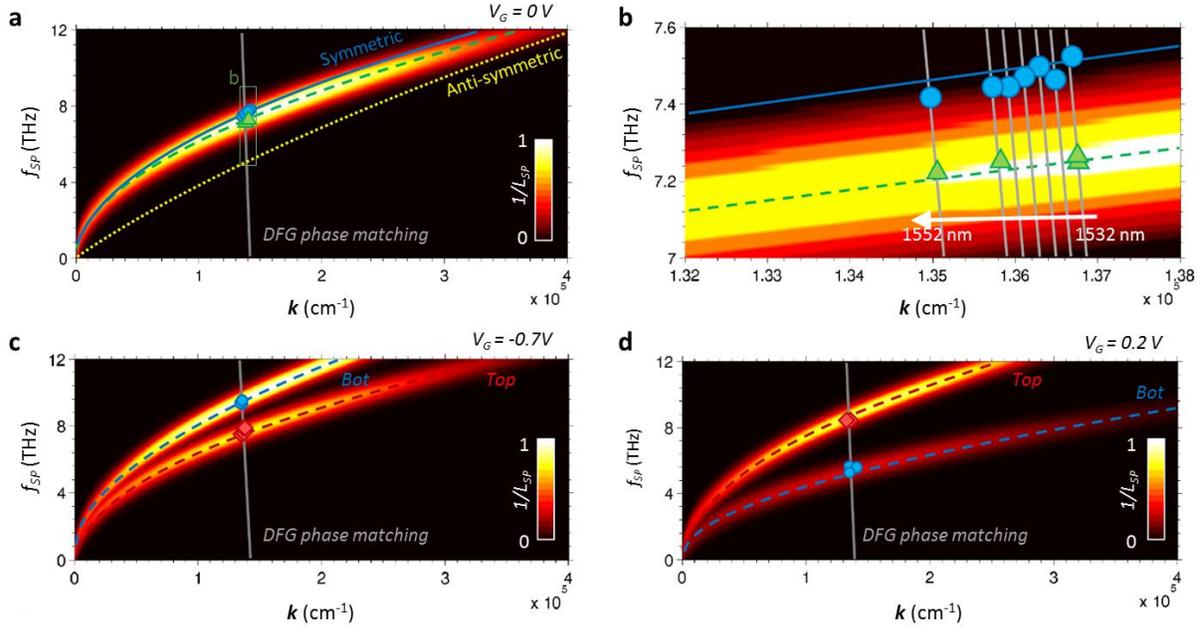

**Figure 3 | Counter-pumped phase matching conditions. a,** Counter-pumped phase matching condition meets the graphene plasmonic dispersion, for $V_G = 0$ V. Here the green dashed curve shows the calculated dispersion of each graphene atomic layer. The blue solid curve and the yellow dotted curve shows the symmetric and the anti-symmetric mode dispersions when interlayer plasmonic coupling is taken into consideration, and the top-down grey solid line shows the DFG phase matching condition. Blue dots and green triangles are the measured datapoints from Figure 2 and Figure S5.3 respectively. **b,** Zoom in of **a**. Grey solid lines show the phase matching cases by tuning $\lambda_p$ from 1532 nm to 1552 nm, with $f_{SP}$ changes from 7.5 THz to 7.4 THz. **c** and **d,** Under $V_G = -0.7$ V and 0.2 V respectively, dispersions of the bottom layer graphene (blue dashed curves) and the top layer graphene (red dashed curves) are tuned. Measured DFG frequency on the bottom and top graphene layers are marked as the blue dots and red diamonds respectively. Note that the change in polarity of $V_G$ swaps the bottom and top layer dispersion curves. In all the panels, intensity is normalized.

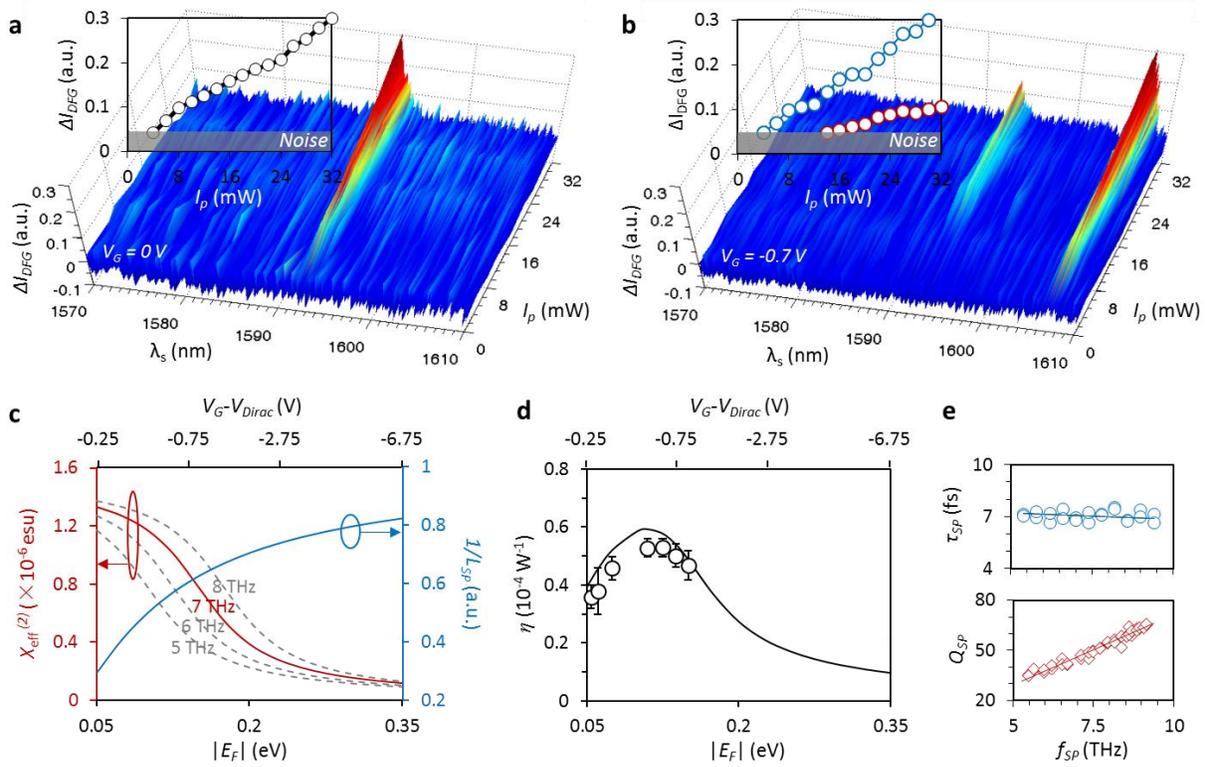

**Figure 4 | Conversion efficiency. a,** Under $V_G$ = 0 V, the measured spectra of the 1593.2 nm enhanced peak are plotted. The peak intensity increases linearly from 0 to 0.3 a.u., with $I_p$ rising from 0 to 32 mW. **b,** Under $V_G$ = -0.7 V, the measured spectra of the enhanced peaks located at 1593.7 nm (bottom layer) and 1607.4 nm (top layer) are plotted. The peak intensities increase linearly as well. Insets in **a** and **b** show the measured $I_p$-$\Delta I_{DFG}$ correlations. **c,** Calculated $\chi_{eff}^{(2)}$ (red curve) and $1/L_{SP}$ (blue curve) during the gate tuning. With increasing $E_F$, graphene offers a lower $\chi_{eff}^{(2)}$ but a higher $1/L_{SP}$. **d,** Overall conversion efficiency $\eta$ of the DFG graphene plasmon, proportional to $\chi_{eff}^{(2)}/L_{SP}$. $\eta$ increases first, and then gradually decreases when $|E_F|$ is higher than 0.13 eV. Here the black dots illustrate the measured results, with a peak value of ≈ 0.6×10$^{-4}$. The error bars are the standard error arising from the intensity noise in the measurement. **e,** Lifetime ($\tau_{SP}$) and Q factor of the generated plasmons – according to the measured $\Delta I_{DFG}$ linewidth.

Supplementary Information of

# Broadband gate-tunable THz plasmons in graphene heterostructures

**S1. Comparison of THz optical sources**
**S2. Theoretical analysis**
S2.1 Dual-layer graphene – optical waveguide interaction
S2.2 Phase matching in the DFG based plasmon generation
S2.3 Plasmonic enhanced $2^{nd}$-order nonlinearity and THz frequency generation
S2.4 Double-layer graphene plasmon: coupling and gating
S2.5 Considerations of DFG versus FWM
**S3. Fabricating the gated dual-layer-graphene - nitride waveguide for THz plasmons**
S3.1 Fabrication process flow
S3.2 Tuning the graphene – nitride DFG-plasmon interactions in the dual-layer graphene structures
**S4. Experimental architecture**
S4.1 Experimental setup
S4.2 Pulsed pump and its modulation
S4.3 CW signal light balanced detection and locked in amplification
**S5. Additional and supporting measurements**
S5.1 Transmission of the GSiNW
S5.2 Pre-saturation of the GSiNW by using CW signal
S5.3 DFG enhanced signal of a GSiNW with 60 nm thick $Al_2O_3$
S5.4 Measurement of the plasmons with pump frequency tuning

**S1. Comparison of THz sources**

Motivated by graphene's unique tunability, long-lived collective excitation and its extreme light confinement, we find an attractive potential of graphene plasmons to realize tunable THz sources [S1]. Table S1 compares some typical THz sources (0.1~50 THz) reported previously [S2-S11]. The unique advantage of the THz plasmon generation in graphene heterostructures is its wide tunability, approximately 10 times higher than the state-of-art tunable QCLs and undulators.

**Table S1 | Comparison of the THz optical sources**

| Type | Output frequency (THz) | Tunability (THz) | Tuning frequency via | Potential to be fast | Ref. |
|---|---|---|---|---|---|
| **Conventional QCLs** | 20 | n/a | | | [S2-S3] |
| **DFG based on crystals** | 1.4 | ≈ 4.7 (4.7 to 9.4 THz) | changing seed frequency | No | [S4] |
| **Frequency comb based on microresonator** | 1.61 | n/a | | | [S5] |
| **Frequency comb + QCL** | 2.5 | n/a | | | [S6] |
| **QCL + DFG** | 34 | ≈ 0.74 (33.72 to 34.46 THz) | tuning temperature | No | [S7] |
| **QCL + gratings** | 38.4 | ≈ 0.60 (38.07 to 38.67 THz) | tuning temperature | No | [S8] |
| **QCL + DBR + DFB** | 3.8 | ≈ 0.6 (3.4 to 4 THz) | modulating static bias (temperature) | No | [S9] |
| **Cherenkov DFG** | 3 | ≈ 3.6 (1.7 to 5.3 THz) | rotating the diffraction grating | No | [S10] |
| **Helical undulator** | 0.1, 0.2, 0.4 | not continuous | changing the diameter of nanowire | No | [S11] |
| **Graphene plasmonic heterostructure** | 7 | ≈ 4.7 (4.7 to 9.4 THz) limited by EDFA | tuning gate voltage | Yes | This work |

**Table S2 | Comparison of the graphene plasmon generation and control.**

| Device design | Excitation Scheme | Excitation Wavelength | Observation method | Gate tunability | Ref. |
|---|---|---|---|---|---|
| **Monolayer graphene sample** | Out-of-plane | Mid-infrared (9.7 to 11.2 μm) pump | $s$-SNOM | Yes | [S12-S13] |
| **Monolayer graphene encapsulated with $h$-BN sample** | Out-of-plane | Mid-infrared (ultrafast; ≈ 200fs) probe – infrared pump | $s$-SNOM | Yes | [S14] |
| **Monolayer graphene nano-antenna sample** | Out-of-plane | Mid-infrared (10.2 to 11.1 μm) illumination | $s$-SNOM | Yes | [S16] |
| **Monolayer graphene nanoresonator sample** | Out-of-plane | Mid-infrared (10 to 12) μm pump | $s$-SNOM | No | [S18] |
| **Monolayer graphene nano-island sample** | Out-of-plane SHG/THG | Mid-infrared source | Mid-infrared optical spectroscopy | Potential | [S17] |
| **Monolayer graphene sample** | Out-of-plane DFG | All-visible wavelength 1-kHz 100-fs mJ pump-probe | mJ pump-probe | No | [S20] |
| **Monolayer graphene nanoribbon sample** | Out-of-plane | Fourier transform infrared broadband source | Fourier transform spectroscopy | Yes | [S15] |
| **Monolayer graphene nanoribbon sample** | Out-of-plane | Broadband infrared source | Fourier transform spectroscopy | No | [S19] |
| **Dual-layer heterogeneous graphene; integrated with on-chip waveguides** | In-plane counter-pumped DFG | All near-infrared pulsed pump (C band) + CW probe (L band) | All near-infrared optical spectroscopy | Yes | this work |

Table S2 compares graphene plasmon generation, observation and control in this work with the state-of-literature techniques [S12-S20]. Figure S1 maps the performances of the state-of-art gate tunable graphene plasmons reported recently. Figure S1a shows, by using ultrathin $Al_2O_3$ dielectric barrier between dual layer graphene, we achieve a full octave tunability, for the first time. Under single volt gating, Fermi level of the graphene atomic layers in our GSiNW can be modulated across the Dirac point. Figure S1b shows, the efficiency of gate tunability in this work is near 1 order higher than the published state-of-art works. Figure 1c maps that this work is unique using 'C+L' optical sources, which is cheap and widely applied in optical systems. Figure 1d highlights that compared to other graphene plasmon generations based optical nonlinearities, the on-chip waveguide design with $\approx 1$ μm$^2$ mode field area enables this work without OPA or femtosecond pump, we apply a ps pulsed pump with 40 mW maximum average power (200 W peak power), the on-chip peak power density can reach 10 GW/cm$^2$, which is $\approx 1$ order higher than previous reports based on out-of-plane implement. The GSiNW design enables the nonlinear process higher efficiency.

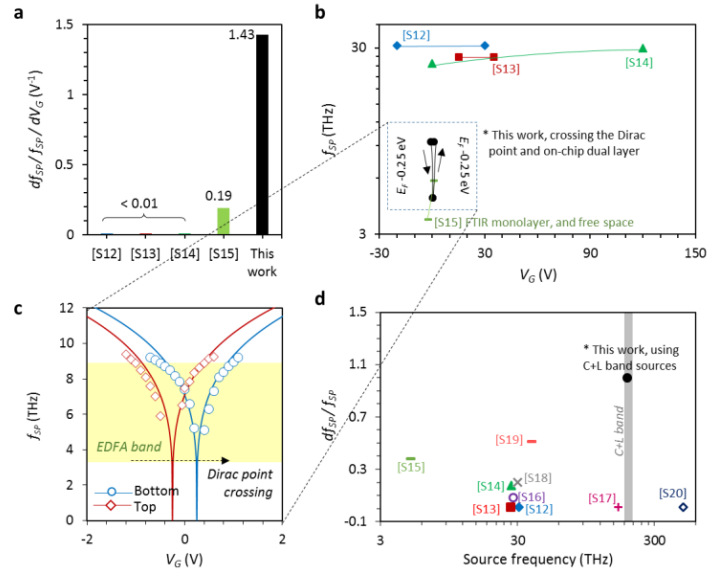

**Figure S1 | Comparison on the gate-tunable THz plasmons**. **a,** Comparison of gate-tuning efficiency [d$f_{sp}$/$f_{sp}$/d$V_G$] for different studies. **b,** Comparison map of $f_{SP}$ with $V_G$ for different studies. **c,** Zoom-in of our $f_{SP}$ versus $V_G$ for the top and bottom graphene layers. **d,** Comparison map of [d$f_{sp}$/$f_{sp}$] versus source frequency across a number of studies.

## S2. Theoretical analysis

### *S2.1 Dual layer graphene – optical waveguide interaction*

Figures S2.1a and S2.1b show the cross-sectional views of the dual-layer graphene nitride and the original nitride waveguide. Figure S2.1c shows the computed effective index dispersion of the TM fundamental mode in the graphene based silicon nitride waveguide, calculated via finite-element method with COMSOL commercial software. Here the index of silicon nitride material ranges from 1.9886 to 1.9904 (188 THz to 200 THz) [S21], the index of $SiO_2$ cladding is fixed at 1.4462, and the index of air is 1. To model the phase matching in dual-layer graphene structure, the effective graphene and pump-signal mode indices need to be examined. For the graphene-nitride structure, the fields of the TM fundamental mode transmitting along

a conventional waveguide [S22] can be written as

$$B_x = \begin{cases} B_1 e^{ik_z t} \cos\left(\frac{k_2 h_{core}}{2} - \varphi\right) \exp k_1\left(\frac{h_{core}}{2} - y\right), & y > \frac{h_{core}}{2} \\ B_2 e^{ik_z t} \cos(k_2 y - \varphi), & -\frac{h_{core}}{2} < y < \frac{h_{core}}{2} \\ B_3 e^{ik_z t} \cos\left(\frac{k_2 h_{core}}{2} + \varphi\right) \exp k_3\left(\frac{h_{core}}{2} + y\right), & y < \frac{h_{core}}{2} \end{cases} \quad (S1)$$

$$E_y = \begin{cases} \frac{ick_1 B_1}{\omega \epsilon_1} e^{ik_z t} \cos\left(\frac{k_2 h_{core}}{2} - \varphi\right) \exp k_1\left(\frac{h_{core}}{2} - y\right), & y > \frac{h_{core}}{2} \\ \frac{ick_2 B_2}{\omega \epsilon_2} e^{ik_z t} \cos(k_2 y - \varphi), & -\frac{h_{core}}{2} < y < \frac{h_{core}}{2} \\ -\frac{ick_3 B_3}{\omega \epsilon_3} e^{ik_z t} \cos\left(\frac{k_2 h_{core}}{2} + \varphi\right) \exp k_3\left(\frac{h_{core}}{2} + y\right), & y < \frac{h_{core}}{2} \end{cases} \quad (S2)$$

$B_{1,2,3}$ are the maximum magnetic field intensities in air, core and $SiO_2$ under the core, and $k_2^2 + k_z^2 = \epsilon_2(\omega/c)^2$, $-k_1^2 + k_z^2 = \epsilon_1(\omega/c)^2$, $-k_3^2 + k_z^2 = \epsilon_3(\omega/c)^2$ are the propagation constants. $\epsilon_1, \epsilon_2, \epsilon_3$ are the permittivities of the zone 1, 2, and 3, respectively. Here $\epsilon_{1,2,3} = n_{1,2,3}^2$, $\varphi$ is the phase constant, and $\omega = 2\pi f$ is the frequency. For the propagation mode, the propagation constant $k_y$ satisfies

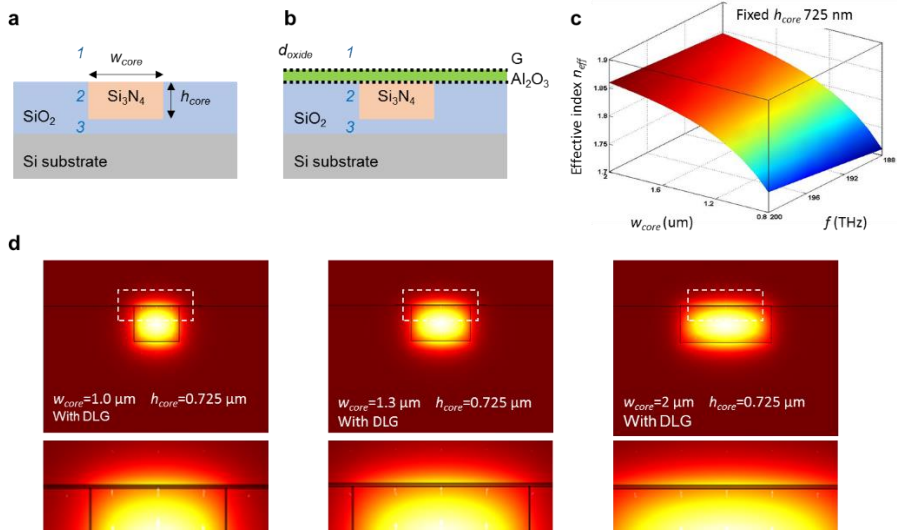

**Figure S2.1 | Mode distributions. a,** Cross-section of the original silicon nitride waveguide without graphene coverage. **b,** Cross-section of the silicon nitride waveguide with graphene-$Al_2O_3$-graphene coverage. **c,** Effective index dispersion of the fundamental TM mode in the silicon nitride waveguide, meshed in the waveguide width ($w_{core}$) and guided frequency ($f$) map. **d,** Simulated $E$-field distributions of the fundamental TM modes in the GSiNW with $w_{core}$ = 1 μm, 1.3 μm and 2 μm. Here the graphene layers are assumed with a $|E_F|$ = 0.1 eV. DLG: dual-layer graphene.

$$\tan(k_y h_{core}) = \left(\frac{k_y}{\epsilon_2}\frac{k_1}{\epsilon_1} + \frac{k_y}{\epsilon_2}\frac{k_3}{\epsilon_3}\right)\left(\frac{k_y}{\epsilon_2}\frac{k_y}{\epsilon_2} - \frac{k_1}{\epsilon_1}\frac{k_3}{\epsilon_3}\right) \tag{S3}$$

$$\tan(2\varphi) = \left(\frac{k_y}{\epsilon_2}\frac{k_3}{\epsilon_3} - \frac{k_y}{\epsilon_2}\frac{k_1}{\epsilon_1}\right)\left(\frac{k_y}{\epsilon_2}\frac{k_y}{\epsilon_2} + \frac{k_1}{\epsilon_1}\frac{k_3}{\epsilon_3}\right) \tag{S4}$$

Because graphene is of considerable index $n_g$ and conductivity $\sigma_g$, it can dramatically modify the boundary conditions. Referring the electromagnetic boundary conditions on the dual layer graphene layer

$$\epsilon_1 E_1 - \epsilon_2 E_2 = \rho_g, B_1 - B_2 = \sigma_g E_2 \tag{S5}$$

Here $\rho_g > 0$ and $\sigma_g > 0$ are the surface charge and conductivity of the layer. For $z \to \infty$, the simulated $E$-field distributions for the GSiNW are shown in Figure S2.1d. Light interacts with graphene layers via the evanescent field. Here graphene-$Al_2O_3$-graphene layer is of 0.4 nm + 30 nm + 0.4 nm thickness.

### S2.2 Phase matching in the DFG based plasmon generation

In the DFG based plasmon generation, energy converts from a pump photon ($f_p$ in C band) to a signal photon ($f_s$ in C band) and a plasmon ($f_{SP}$ in THz band). During this process, momentum is conserved. Thus we write the energy matching and phase matching condition as

$$hf_s + hf_{SP} = hf_p, \vec{k_s} + \vec{k_{SP}} = \vec{k_p} \tag{S6}$$

Here $h$ is the Planck constant, $k_s$, $k_{spp}$ and $k_p$ are the wavevectors of the signal, plasmon and pump. In optics, $k = 2\pi/\lambda = 2\pi n_{eff}/cT = 2\pi f n_{eff}/c$, where $n_{eff}$ is the effective index and $c$ is the light speed in vacuum. With the counter-propagation pump-signal geometry, we rewrite Eq. (S6) to be

$$\begin{cases} f_s n_s - f_{SP} n_{SP} = -f_p n_p \\ f_s + f_{SP} = f_p \end{cases} \tag{S7}$$

Here $n_p$, $n_s$, and $n_{SP}$ are the effective indexes of the pump light, signal light, and the plasmon respectively. To satisfy the phase matching, $f_p$, $f_s$ and $n_{SP}$ should be selected and adjusted carefully:

$$\frac{f_s}{f_p} = \frac{n_{SP} - n_p}{n_{SP} + n_s}, n_{SP} > n_p \tag{S8}$$

In our measurements of the main text $f_p$ is fixed at 195.8 THz (1531.9 nm) (in Supplementary Section S4.5, the pump wavelength is varied). The effective index $n_p$ of the silicon nitride waveguide at $f_p$ is ≈ 1.77 (waveguide with $w_{core}$ = 1.3 μm and $h_{core}$ = 0.75 μm). In our measurements, $f_s$ is scanned from 192.3 THz to 177.5 THz (1560 nm to 1690 nm); the effective index $n_s$ ranges from ≈ 1.77 to 1.75. The material index of graphene $n_g = n_{g,r} + i n_{g,i}$ plays the key role in this equation. One can derive $n_g$ from $\sigma_g$, as

$$\sigma_g(f, E_F, \tau, T) = \frac{ie^2(2\pi f - i/\tau)}{\pi \hbar^2}\left\{\frac{1}{\left(2\pi f + \frac{i}{\tau}\right)^2}\int_0^\infty \epsilon\left[\frac{\partial f_d(\epsilon)}{\partial \epsilon} - \frac{\partial f_d(-\epsilon)}{\partial \epsilon}\right]d\epsilon - \int_0^\infty \left[\frac{f_d(-\epsilon) - f_d(\epsilon)}{(2\pi f + i/\tau)^2 - 4(\epsilon/\hbar)^2}\right]d\epsilon\right\} \tag{S9}$$

$$\sigma_{g,intra} = \frac{ie^2 E_F}{\pi \hbar (2\pi f + \frac{i}{\tau})} \tag{S10}$$

$$\sigma_{g,inter} = \frac{ie^2 E_F}{4\pi\hbar} \ln\left[\frac{2|E_F|-\hbar(2\pi f+\frac{i}{\tau})}{2|E_F|+\hbar(2\pi f+\frac{i}{\tau})}\right] \tag{S11}$$

Hence,

$$\epsilon_g = \frac{-\sigma_{g,i}+i\sigma_{g,r}}{2\pi f \Delta} \tag{S12}$$

$$(n_{g,r}+in_{g,i})^2 = \epsilon_{g,r}+i\epsilon_{g,i} \tag{S13}$$

$$n_{g,r} = sqrt\left(\frac{-\epsilon_{g,r}+\sqrt{\epsilon_{g,r}^2-\epsilon_{g,i}^2}}{2}\right), n_{g,i} = sqrt\left(\frac{-\epsilon_{g,r}+\sqrt{\epsilon_{g,r}^2-\epsilon_{g,i}^2}}{2}\right) \tag{S14}$$

In above equations, $E_F$ is the Fermi level, $\tau = 10^{-13}$ s is the relaxation lifetime, $T$ is the temperature, $f_d(\epsilon) = \{exp[(\epsilon-E_F)/k_B T]+1\}^{-1}$ is the Fermi-Dirac distribution, $\hbar = 1.05\times10^{-34}$ eV·s is the reduced Planck constant, $k_B = 1.3806505\times10^{-23}$ J/K is the Boltzmann's constant, and $e = -1.6\times10^{-19}$ C is the unit charge. When graphene is gated, $n_g$ is much higher than $n_p$ or $n_s$ [S23], corresponding to the $f_{sp}$ much smaller than $f_p$ or $f_s$. Figure S2.2.1a and Fig. S2.2.1b shows the calculated conductivity and the permittivity of graphene using the Kubo formalism [S24-S26]. When $\sigma_{g,i} > 0$, graphene can support surface plasmons.

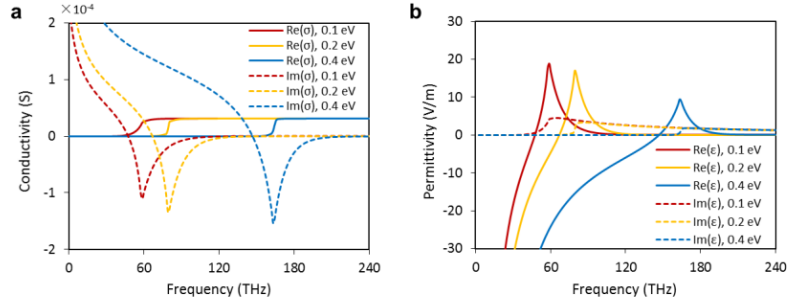

**Figure S2.2.1 | Graphene conductivity and permittivity. a,** Calculated conductivity of graphene, with Fermi level at 0.1 eV (red), 0.2 eV (yellow), and 0.4 eV (blue). **b,** Calculated permittivity of graphene, with Fermi level at 0.1 eV (red), 0.2 eV (yellow), and 0.4 eV (blue). Here the solid curves show the real parts while dashed curves show the imaginary parts.

When the plasmon frequency is lower than Landau damping regime, we get the momentum-frequency ($k_{SP}$-$f$) dispersion of graphene. With the boundary conditions, it could be approximately simplified as a quadratic function (see also Eq. S28-S29) as

$$k_{SP} = Af^2 \tag{S15}$$

$$A = \frac{(1+n_{SP}^2)\hbar\pi^2}{2\alpha c v_F} \tag{S16}$$

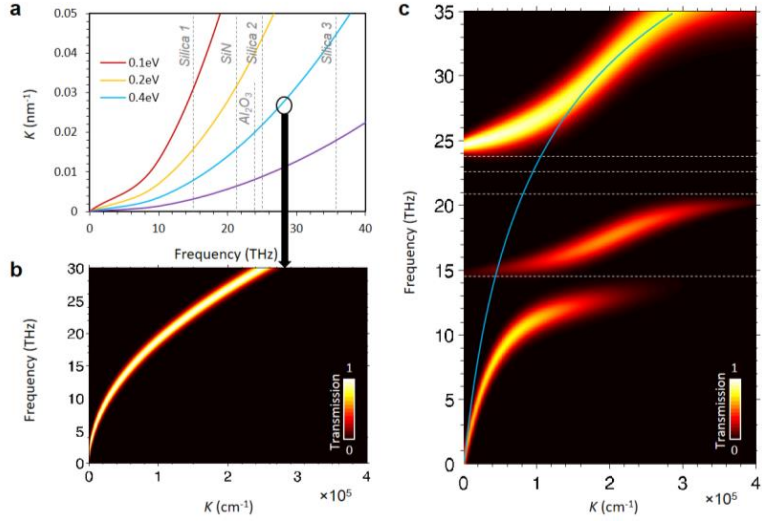

**Figure S2.2.2 | Phase matching. a,** Graphene *f-k* dispersion, under Fermi level of 0.1 eV, 0.2 eV, 0.4 eV and 0.8 eV. **b,** Calculated $1/L_{SP}(k,f)$ of graphene with Fermi level 0.4 eV, based on RPA method. **c,** $1/L_{SP}(k,f)$ map describing the plasmon-phonon couplings in the GSiNW. Phonon frequencies are marked out with the dashed lines. In **b** and **c**, the values of the transmission ~ $1/L_{SP}(k,f)$ is normalized to be 1.

Here $\alpha$ is the fine structure constant, and $E_F$ is the Fermi level of graphene. With the relationship of $n_g$ and $E_F$, the calculated $k_{SP}$-$f$ curves of graphene with Fermi level ranging from 0.1 eV to 0.8 eV are shown in Figure S1.2.2a. In Figure S1.2.2a, the grey lines show the phonon resonance locations of the $Si_3N_4$ and $SiO_2$. $f_{silica,1}$ = 14.55 THz, $f_{silica,2}$ = 24.18 THz, $f_{silica,3}$ = 36.87 THz, $f_{SiN}$ = 21.89 THz, and $f_{Al2O3}$ = 22.4 THz [S27-S33].

Furthermore, by using random phase approximation (RPA) method [S30-S33], we calculate the plasmon coupling based loss $L_{SP}(k,f)$ along the GSiNW, with consideration of the phonon couplings.

$$L_{SP}(\boldsymbol{k},f) = -Im\left\{1 - \frac{e^2}{2k\epsilon_1}\Pi_0(\boldsymbol{k},f) - \sum_j f_{ph,j}\Pi_0(\boldsymbol{k},f)\right\} \tag{S17}$$

$$\Pi_0(\boldsymbol{k},f) = -\frac{g_s}{4\pi^2}\sum \int \frac{f_d(\epsilon_s) - f_d(\epsilon_{sk})}{2\pi f\hbar + \frac{i\hbar}{\tau} + \epsilon_s - \epsilon_{sk}} d\boldsymbol{k} F(s,\boldsymbol{k}) \tag{S18}$$

Here $f_{ph,j}$ is the phonon resonances, $g_s = 4$, $f_d(\epsilon)$ the Fermi-Dirac distribution, $\epsilon_s = sv_F$, $\epsilon_{sk} = sv_F k$, $s = \pm 1$, $F(s,\boldsymbol{k})$ is the band overlap function of Dirac spectrum, which equals 1 for the waveguide geometry. Figure S2.2.2b provides the simulated $1/L_{SP}(k,f)$ map of graphene for the Fermi level at 0.4 eV, without considering the plasmon-phonon couplings. Figure S2.2.2c shows the RPA map with consideration of the phonon couplings.

### S2.3 Plasmon enhanced 2nd-order nonlinearity and the THz frequency generation

Graphene is a single atomic layer with honeycomb structure, therefore, second-order nonlinear effects are described by the second-order surface nonlinear conductivity [S34]. With light transmitting along graphene with a wavevector $k$ parallel to the 2D layer plane, the second-order nonlinear polarizability $\chi^{(2)}$ can be large [S35]. In graphene, an effective $\chi^{(2)}$ can be written as

$$\frac{\partial^2 \chi_{ijk}^{(2)}}{\partial k^2} = \frac{e^2}{4\pi^4 \hbar^2 f_p f_s} \left\{ \left[ \frac{f(k_1)-f(k_3)}{\omega_{31}-2\pi f_p-i\gamma} + \frac{f(k_1)-f(k_2)}{2\pi f_s-\omega_{21}-i\gamma} \right] \frac{\mu_{32}^i v_{31}^i v_{21}^k}{\omega_{32}-2\pi f-i\gamma} - \left[ \frac{f(k_1)-f(k_3)}{\omega_{31}-2\pi f_p-i\gamma} + \frac{f(k_2)-f(k_3)}{2\pi f_s-\omega_{23}-i\gamma} \right] \frac{\mu_{21}^i v_{31}^i v_{32}^k}{\omega_{21}-2\pi f-i\gamma} \right\} \quad (S19)$$

$$\mu_{ab} = \frac{iev_F}{\omega_{ab}} <a|\sigma_g|b>, v_{ab} = v_F <a|\sigma_g|b>, \omega_{ab} = \frac{E(k_a)-E(k_b)}{\hbar} \quad (S20)$$

Here $f(k)$ is the occupation number state $k$, $k_1$, $k_1$ and $k_3$ satisfy $k_1+ k_p= k_3$, $k_1+ k_s= k_2$. Here $\sigma_g$ is the 2D Pauli matrix vector, $<a>$ and $<b>$ are the states, $\gamma$ is the scattering rate, and $v_F=E_F/(\hbar k)$ is the Fermi velocity. By approximating $k_BT \to 0$, $2\pi f >> v_F k$, $f_p \approx f_s$, along the graphene, the second-order nonlinear polarizability $\chi_{ijk}^{(2)}$ can be simplified as

$$\chi_{eff}^{(2)} = \frac{e^3}{4\pi^2 \hbar^2} \frac{1}{k\sqrt{f_s f_p}} \left[\frac{\pi}{2} + \arctan\left(\frac{2\pi\sqrt{f_s f_p}-2v_F k_F}{\gamma}\right)\right] \quad (S21)$$

Here $k_F = (2m_e E_F)^{1/2}$ is the Fermi momentum. The simulated $\chi_{eff}^{(2)}$ is shown in Figure S2.3a: A higher $E_F$ brings a lower $\chi_{eff}^{(2)}$. Here $f_p$ is fixed as 1.93 THz. Hence we write the $E$-field intensity of the generated plasmon as

$$E_{SP} = \frac{\chi_{eff}^{(2)} E_{p,y}(\frac{h_{core}}{2}) E_{s,y}(\frac{h_{core}}{2})}{L_{SP}} \quad (S22)$$

Here $L_{SP}$ is shown in Eq. (S15). Referring to Eq. (S1) and Eq. (S2), here $E_{p,y}$ $E_{s,y}$ are the $E$-field of the pump and the signal respectively, with $z = ct/n_g$, $k_p = 2\pi f_p n_{eff,p}/c$, $k_s = 2\pi f_s n_{eff,s}/c$. The real part of Eq. (S22) can be approximately simplified as

$$E_{SP}(t) = \frac{1}{2} A_{SP}(t) \cos\left[2\pi \frac{(f_p n_p+f_s n_s)}{n_{SP}} t\right] + \frac{1}{2} A_{SP}(t) \cos\left[2\pi \frac{(f_p n_p-f_s n_s)}{n_{SP}} t\right] \quad (S23)$$

$$A_{SP}(t) = \frac{A_p(t) A_s(t) \chi_{eff}^{(2)}}{L_{SP}} \exp(-n_{g,i}) \quad (S24)$$

$A_{SP}(t)$ determines the loss of the surface plasmon wave. Here $A_p(t)$ and $A_s(t)$ are the amplitudes of the pump and signal respectively. In this equation, we get the frequency of the surface plasmon $f_{SP} = (f_p n_p + f_s n_s)/n_{SP}$, and the frequency of heterodyne beat $f_B = (f_p n_p + f_s n_s)/n_{SP}$. Referring to the DFG energy balance $f_{SP} = f_p - f_s$, the effective index of the plasmon $n_{SP}$ satisfies

$$n_{SP} = \frac{f_p n_p+f_s n_s}{f_p-f_s} \quad (S25)$$

This equation corresponds to the Eq. (S8) perfectly. For $f_p$ and $f_s$ located in 'C+L' optical communications band and with $f_{SP} \approx 8$ THz, the $n_{SP}$ satisfying the phase matching condition could be approximately calculated to be $\approx 80$. $n_{SP}$ is also determined by $f_{SP}$ and the Fermi level $E_F$, from Eq. (S15). Figure S2.3b plots the graphene dispersion $n_{SP}$ $(f_p, f_s)$ at $E_F = 0.1$ eV, and the DFG phase matching $n_{SP}$ from Eq. (S25) together. This figure shows that the DFG based graphene plasmon generation is related to $E_F, f_p, f_s$ and the waveguide structure, concisely together in one figure.

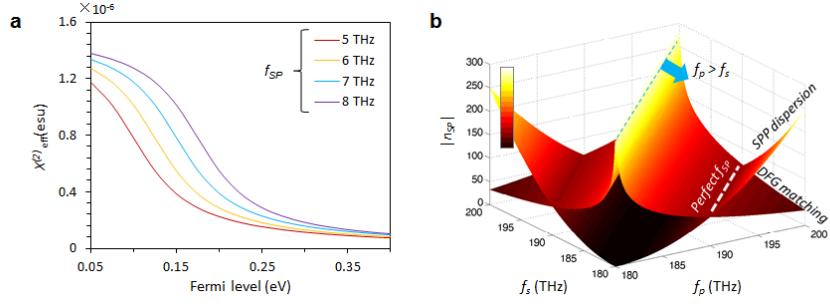

**Figure S2.3 | 2nd-order nonlinear polarizability and effective indices of the surface plasmons. a,** Calculated curves of $\chi^{(2)}_{eff}$ with $f_{SP}$ at 5 THz (red), 6 THz (yellow), 7 THz (blue), and 8 THz (purple). **b,** For $w_{core}$ = 1 μm and $h_{core}$ = 725 nm, to satisfy the phase matching, $n_{SP}$ is determined by both $f_p$ and $f_s$.

## S2.4 Dual-layer graphene plasmon: coupling and gating

In Sections S2.2 and S2.3, phase matching of DFG based on monolayer graphene is analyzed, without considering the possible plasmon coupling of the separated graphene layers in the grphene-Al$_2$O$_3$-graphene system. When the distance between the two graphene layers is small enough, the graphene-Al$_2$O$_3$-graphene could be regarded as a topological insulator-like system in which plasmonic mode coupling can occur [S35-S39]. We schematically show the dual-layer graphene structure in Figure S2.4a. Here the thickness of Al$_2$O$_3$ is taken into consideration as *d*. Compared to single layer graphene, the situation in the gated graphene-Al$_2$O$_3$-graphene structure could be regarded as a capacitor: when stable, the top layer graphene charges +*Q*, and the bottom layer graphene charges -*Q*. The Fermi level of a monolayer graphene is written as [S40]

$$E_F = \hbar|v_F|\sqrt{\pi n} \tag{S26}$$

We note that the initial Fermi levels of the top and the bottom layer graphene would be different ($E_T$, $E_B$): once the graphene-Al$_2$O$_3$-graphene capacitor formed, the top layer graphene is positively charged while the bottom layer is negatively charged, $n_{electron}=n_{hole}=Q/eS_g=C_GV_G/eS_g$. That means, the carrier densities of the top layer ($n_T$) graphene and the bottom layer graphene ($n_B$) would be different. Considering CVD graphene in air is *p*-doped initially, $E_T > E_B$ when V$_G$ > 0. Assuming the top and bottom layer graphene has the same size $S_g$ = 80×20 μm$^2$, a capacitance $C_G$=2×10$^{-7}$ F/cm$^2$, and the initial Fermi levels (before gating) $E_{T0} = E_{B0}$ = - 50 meV ($n_{hole\_0} \sim$ 2×10$^{11}$/cm$^2$), Figure S2.4b shows the resulting computed $V_G$-$E_F$ correlation. Since $E_F$ determines the dispersion of graphene plasmon, Figure S2.4b predicts that two plasmons with different $f_{SP}$ could be generated simultaneously and tuned differently with gate voltage in the graphene-Al$_2$O$_3$-graphene system. Figure S2.4c simulates the dispersions of the dual-layer graphene, without interlayer coupling. Figure S2.4d simulates the dispersions of top and bottom graphene surface plasmons, with reference to the initial one, $E_{T0} = E_{B0}$ = - 50 meV at $V_G$ = 0 V.

However, when the interlayer coupling distance *d* is small enough, the two independent plasmon modes would couple with each other to form two hybrid modes, symmetric and antisymmetric. In the low frequency regime, the dispersions of the symmetric and antisymmetric modes are described as:

$$f_{sym} = \frac{1}{2\pi}[\frac{2e^2}{\varepsilon}(E_T + E_B)k_{sp}]^{1/2} \tag{S28}$$

$$f_{asym} = \frac{1}{2\pi}[\frac{4e^2 E_T E_B d}{\varepsilon(E_T+E_B)}]^{1/2} k_{sp} \tag{S29}$$

Here $\varepsilon$ is the background permittivity and $d$ is the dielectric layer thickness. In Figures S2.4e and S2.4f, we show the calculated dispersions of $f_{op}$ ($k_{sp}$) and $f_{ac}$($k_{sp}$), with $e^2/\varepsilon \approx 7\times10^5$ THz²nm·eV⁻¹, and $d$ = 30 nm. Plasmon coupling further splits the dispersion curves in THz region.

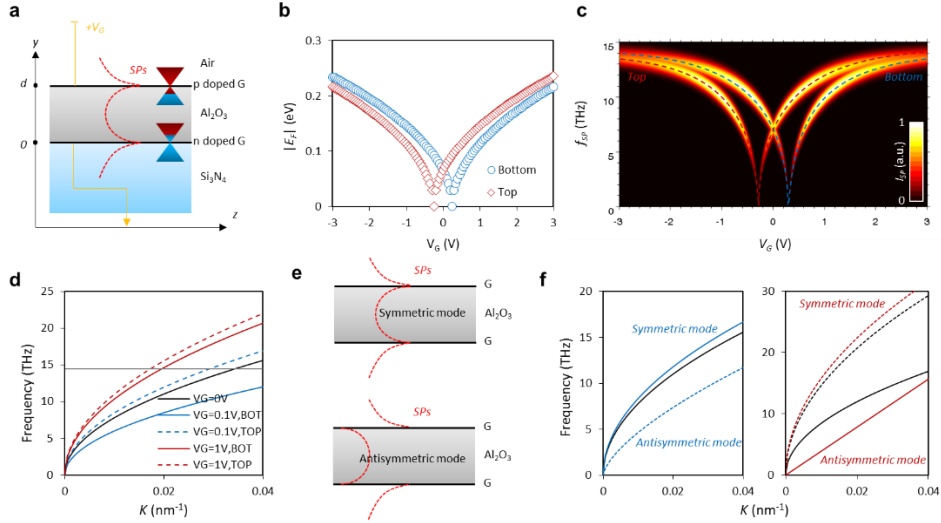

**Figure S2.4 | Graphene-Al₂O₃-graphene system. a,** Schematic configuration. **b,** Correlation of gate voltage and Fermi level. Blue circles denote the bottom layer and red diamonds denote the top layer. **c,** Simulated plasmon dispersions on the top and bottom graphene layers. When $V_G$ = 0 V, the two graphene layers have the same $E_F$ of ≈ 50 meV intrinsically. **d,** Dispersions under different $V_G$, without interlayer coupling. Solid curves denote the independent bottom layer and dashed curves denote the independent top layer. **e,** Symmetric and antisymmetric modes. **f,** Dispersions of coupled modes, which are mode-split from the original ones. Black curves: independent modes (solid: bottom layer; dashed: top layer). Left panel denotes $V_G$ = 0V and right panel denotes $V_G$ = -1V.

## *S2.5 Considerations of DFG versus FWM*

Graphene also has large $\chi^{(3)}$, which offers third-order optical nonlinearity, e.g. four wave mixing (FWM) [S41]. One might wonder if the enhancement of the signal is plausible from FWM instead of DFG [S42-S45]. Here analysis is shown theoretically to exclude the influence of FWM, in our pump-signal counter-launched configuration. In a typical degenerate FWM process, the photon energy transfers from pump to signal and idler, with energy and momentum matching. When the propagation directions of the pump and the signal are opposite, once FWM occurs we have

$$2f_p = f_s + f_i \tag{S30}$$

$$2\vec{k}_p = -\vec{k}_s + \vec{k}_i \tag{S31}$$

Here $f_p$, $f_s$ and $f_i$ are the frequencies of the pump, signal and idler, $k_p$, $k_s$ and $k_i$ are the momentums of the pump, signal and idler, $k=2\pi f n_{eff}/c$, respectively. The dispersion could be written as

$$2f_p(n_{eff,i} - n_{eff,p}) = f_s(n_{eff,i} + n_{eff,s}) \tag{S32}$$

Here $n_{eff,i}$, $n_{eff,p}$ and $n_{eff,s}$ are the effective mode indexes. To satisfy this equation when the frequency difference of $f_p$ and $f_s$ is smaller than 5 THz, $n_{eff,i}$ would have to be ≈ 3 times larger than $n_{eff,p}$ or $n_{eff,s}$. However, for the FWM-generated mode, its $n_{eff,i}$ cannot be larger than the index of the waveguide core. Hence FWM cannot occur in our counter-launched pump-signal configuration in this case.

## S3. Fabricating the gated dual-layer-graphene - nitride waveguide for THz plasmons

### S3.1 Fabrication process flow

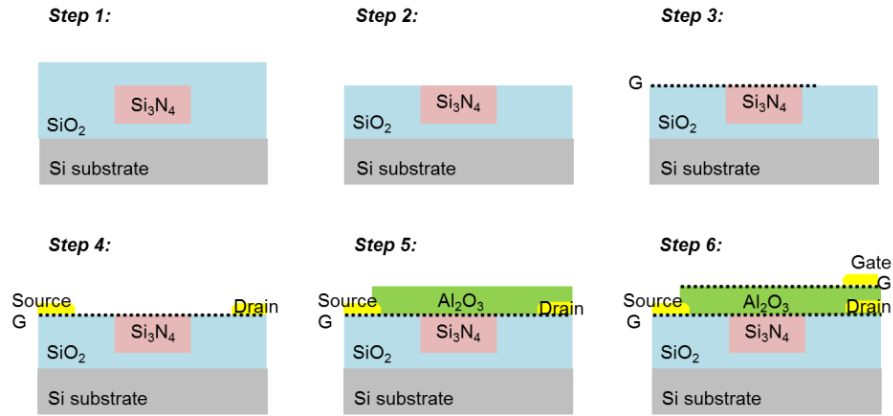

**Figure S3.1 | Nanofabrication process of the dual-layer-graphene nitride plasmon structure.** The graphene layers are transferred onto the nitride waveguide along with the source-drain-gate electrodes and the $Al_2O_3$ dielectric barrier layer deposition.

Figure S3.1 shows the fabrication process steps of graphene on the silicon nitride waveguides (GSiNWs). As shown in *step 1*, the chips are fabricated at the Institute of Microelectronics Singapore, with the silicon nitride waveguide buried in $SiO_2$ cladding. There are 4 straight waveguides in every chip with a width of 1 μm, and length of ≈ 3 mm. The undercladding oxide is 3 μm thick, the height of the waveguide core is 725 nm, and the top oxide cladding is 2.5 μm. The chip is chemically etched by using wet buffered oxide etching (BOE) method in *step 2* (plasma-based dry etching is also available). After etching, the distance between the core and the top oxide surface is less than 20 nm, ensuring the strong light-graphene interaction. In *step 3*, a chemical vapor deposition (CVD) grown monolayer graphene is transferred onto the chip using wet transfer, followed by photolithography patterning and oxygen plasma etching. This graphene layer serves as the bottom layer graphene with a size of 100 μm× 40 μm. Next, the Ti/Au (20/50 nm) contact pad is deposited using electron beam evaporation, working as source-drain electrodes. By using the source and drain, resistance of the bottom layer graphene could be measured. In *step 5*, a thin 30 nm layer of $Al_2O_3$ in deposited using atomic layer deposition (ALD), providing sufficient capacitance for the graphene based semiconductor chip. Finally, as shown in *step 6*, on the top of the $Al_2O_3$ insulator, another graphene layer is transferred, aligned and linked with the gate.

*S3.2 Tuning the graphene – nitride DFG-plasmon interactions in the dual-layer graphene structures*

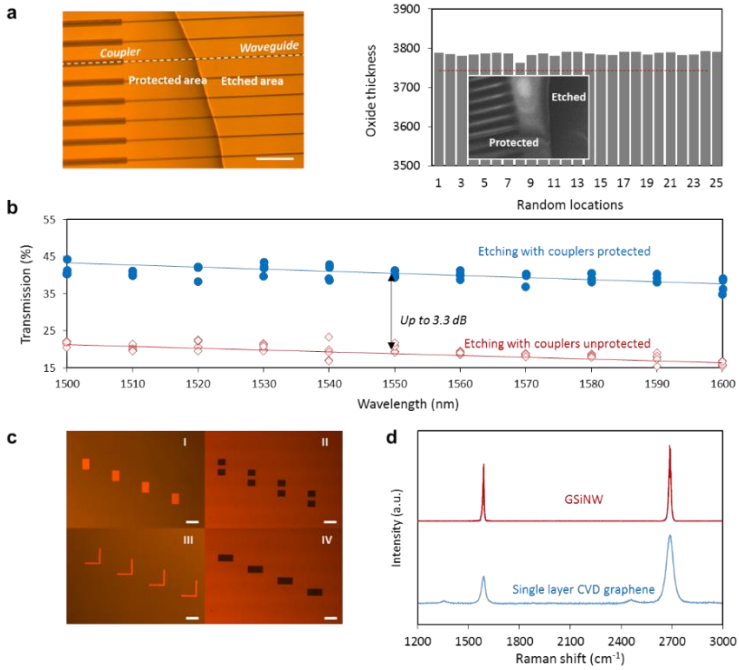

**Figure S3.2 | Chip processing for dual-layer graphene interaction. a,** (Left) Optical micrograph of the processed chip, with arrayed input-output straight waveguides and inverse couplers. Brighter area is etch-controlled down to 100 nm for the graphene-nitride interaction. Scale bar: 100 μm. (Right) Etched thickness. **b,** Average oxide thickness after etching, remaining <50 nm oxide upon the core . **c,** Designed masks for the source-drain-gate implementation. Scale bar: 150 μm. **d,** Normalized graphene Raman spectra, before and after GSiNW transfer and electrodes processing. Blue curve: single layer CVD graphene; red curve, the GSiNW.

Figure S3.2a (left) illustrates a top-view optical micrograph of the etched silicon nitride waveguides (SiNWs). The edge of the etched and the non-etched areas is clear. To reduce the scattering and coupling loss of the etched waveguides, the inverse taper couplers at the input and output facets are carefully protected by photoresist. Figure S3.2a (right) shows the after-etch oxide thickness at random locations, with an uncertainty of ± 10 nm. The oxide thickness refers the distance between the top surface and the bottom Si substrate of a chip. The thickness data is measured by using an optical interferometer at 480 nm

wavelength, with the SiO$_2$ refractive index fixed at 1.4594. Inset is the SEM image focused on the etched edge.

In the experiment, two etching methods were applied: dry etching (via oxygen plasma) and wet etching (via hydrofluoric acid). Figure S3.2b compares the losses of the devices with the same etched depth ≈ 2.5 µm, for different process conditions. It shows that we can get etched chips of acceptable loss (less than 4 dB), via either dry etching or wet etching, but the coupler protection is necessary. Figure S3.2c shows the masks for the graphene-Al$_2$O$_3$-graphene structure fabrication. Patterns marked by I, II, III, IV are for lithography operations of the bottom layer graphene, bottom layer Au electrodes, top layer graphene, and top gate respectively. Figure S3.2d illustrates the resulting graphene Raman spectrum, before and after transfer onto the chip. Pumped with a 514 nm laser and after transfer, the graphene defect $D$ peak is negligible, the $G$ peak width is ≈ 6 cm$^{-1}$, and the $2D$ peak width is ≈ 14. Intensity ratio of $G$ to $2D$ is ≈ 0.75. The Raman spectra are comparable to that of monolayer and dual-layer CVD graphene measured during our fabrication.

## S4. Experimental architecture

### S4.1 Experimental setup

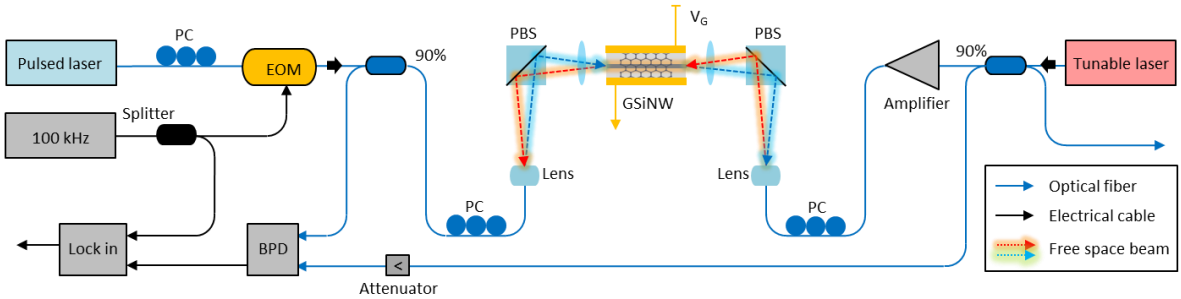

**Figure S4.1 | Experimental setup.** Measurement setup: A mode-locked picosecond laser serves as the ~ 400 pJ pump ($f_p$ = 195.8 THz, 2.2 ps pulse duration, 39.1 MHz repetition rate, and 200 W peak power), which is slowly modulated at 100 kHz sinusoidally for single phase lock-in detection. A broadband tunable CW laser serves as the signal frequency $f_s$, amplified in the 1570 nm to 1610 nm band up to 1.6 W. The plasmon signal is detected in a balanced photodetector, with lock-in detection.

Figure S3.1 illustrates the experimental setup. A mode-locked pump pulse is launched into the GSiNW from the left, while an amplified continuous-wave (CW) signal is counter-launched from the right, both in TM polarizations. In the DFG process, the energy of the converted signal photon arises from the pump photon less the plasmon energy - the generation of the plasmons could thus be observed by monitoring the transmitted signal intensity on the left output. To directly detect the DFG plasmon signal over the noise, we implement a 100 kHz modulation of the mode-locked 39.1 MHz pump laser with lock-in filtering and amplification, along with balanced detection. To enable the detection of DFG plasmon signal, the launched light beams are TM polarized. A high power mode-locked laser is applied the pump and a pre-amplified CW tunable laser is applied as the signal. Balanced photodetection (BPD) and lock-in amplification are implemented to extract the small plasmon signal from white noise exactly and clearly.

*S4.2 Pulsed pump and its modulation*

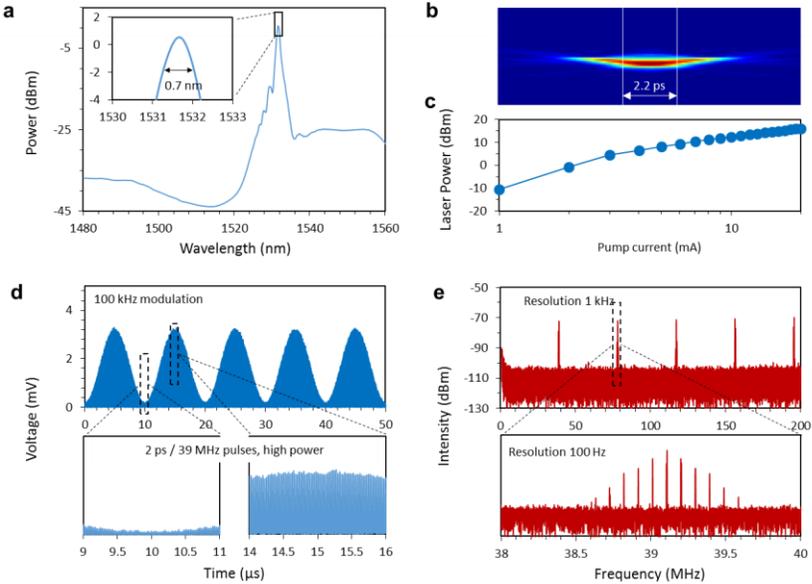

**Figure S4.2 | Pump filtering and modulating. a,** Spectra of the ps pulsed pump, with 0.7 nm spectral linewidth. **b,** Pulsewidth of the pulsed pump, measured by using FROG. **c,** Measured averaged power of the ps pulsed pump. **d,** Temporal profile of the pump, modulated by a 100 kHz sinusoidal signal for lock-in amplification, with the embedded 39.1 MHz pulses inside. **e,** Corresponding electronic spectrum of the modulated pump, 39.1 MHz peaks with 100 kHz harmonics are clear.

Here we use a nonlinear process to detect the in-plane graphene plasmons. To enhance the DFG nonlinear signal detection, a mode-locked picosecond fiber laser serves as the pump, which is pre-filtered and modulated. The spectral and temporal profile of the near-transform-limited pump launched onto the chip is illustrated in Figure S4.2. The spectrum is centered at 1531.8 nm (195.8 THz) with an ≈ 0.7 nm linewidth (Figure S4.2a). Figure S4.2b shows the temporal profile with 2.2 ps full-width half-maximum, measured by frequency-resolved optical gating (FROG), with a maximum average power of 16.1 dBm (40.7 mW) at 1531.9 nm and a quasi-linear increase (Figure S4.2c). Figure S4.2d shows the modulated pulsed pump, with the slow 100 kHz envelope and the embedded 39.1 MHz pulses inside. Figure S4.2e shows the corresponding electronic spectrum.

*S4.3 CW signal light balanced detection and locked-in amplification*

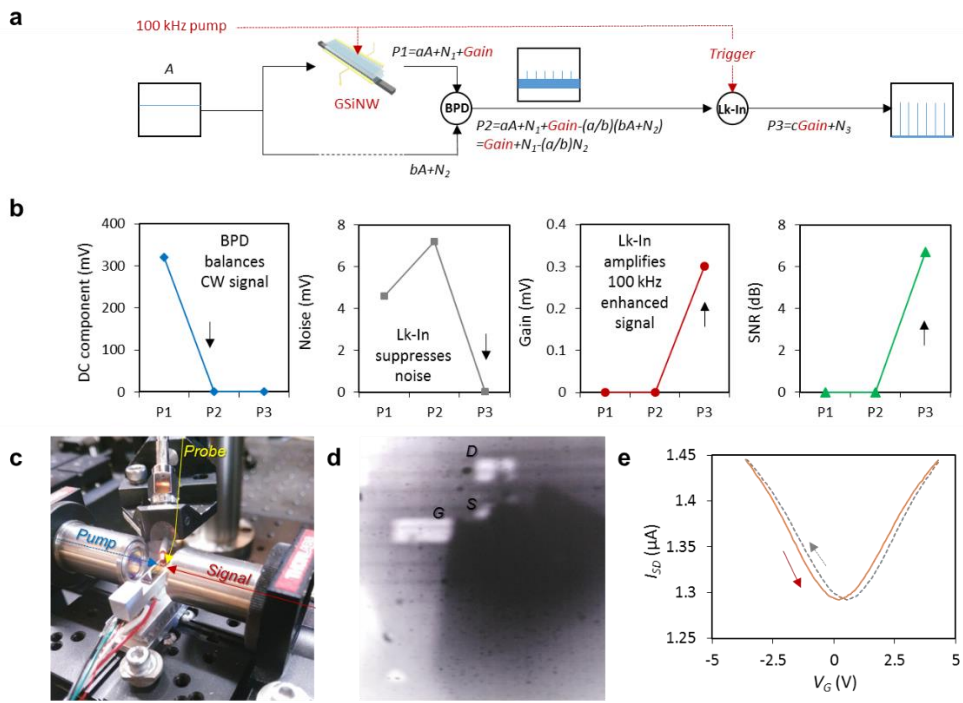

**Figure S4.3 | Signal balanced detection and locked-in amplification. a,** Schematic of the CW signal processing. Lk-In: lock-in amplification. **b,** Comparisons: DC component (original CW light), noise component, 100 kHz gain, and the resulting SNR. Here $P1,2,3$ corresponds to the tap-out points in **a**. **c** and **d,** Gating the graphene-based semiconductor chip. **e,** Hysteresis loop of the GSiNW with $V_G$ from -4 to 4 V (red) and 4 to -4 V (grey).

Figure S4.3a illustrates the setup to detect the weak DFG (with a 100 kHz modulation) from a strong background signal (CW) schematically. A CW tunable laser with intensity *A* is divided to be two paths. One passes the GSiNW while the other one serves as a reference. Then the DFG enhanced path with a 100 kHz gain is balanced by the reference, eliminating the CW component *A*. The dynamic intensity of the balanced signal has both gain and noise components. To extract the gain from noise, we use a lock-in amplifier at 100 kHz clock. Here, $N_{1,2,3}$ denote the noises and *a, b, c* are the attenuation and amplification factors. Correspondingly Figure S4.3b compares the measured intensities of CW signal (the DC component), the noise, the newly generated signal (from the DFG process, with 100 kHz oscillation), and the SNR before the balanced photodetector (BPD) (P1), after the BPD (P2), and after the lock-in amplifier (P3). We demonstrate that the BPD is predominantly used for DC balancing while the lock-in amplifier is applied to lock and amplify the 100 kHz gain. Figures S4.3c and S4.3d show the gating of the chip-scale GSiNWs with the micro-probes. $V_G$ is tuned up to ± 4 V with 10 mV accuracy. Figure S4.3e illustrates the measured hysteresis loop of the GSiNW with $V_{SD}$ at 10 mV. When $V_G \approx 0.25$ V, the bottom layer graphene approaches the Dirac point.

## S5. Additional and supporting measurements

### S5.1 Transmission of the GSiNW

Figure S5.1a shows the chip-scale 1500 to 1600 nm normalized transmitted spectrum, before and after covering with the graphene-Al$_2$O$_3$-graphene hybrid layer. The transmission of the nitride waveguide before etching is normalized as 0 dBm, and the launched power is ≈ 1 mW (significantly lower than the graphene saturated threshold). The initial 3.4 dB insertion loss is from the wet-etch chip processing; graphene coverage subsequently brings additional loss due to its monolayer broadband optical absorption. The loss of the shorter wavelengths is lower (red curve), perhaps due to the better mode field confinement. The graphene induced loss is ≈ 7.3 dB at 1500 nm (0.09 dB/μm), ≈ 8.3 dB at 1550 nm (0.1 dB/μm), and ≈ 9.5 dB at 1600 nm (0.12 dB/μm). Figure S5.1b tables the pump-signal polarization combinations – only when both the pump and signal are of TM polarization can DFG and the resulting THz plasmons be excited.

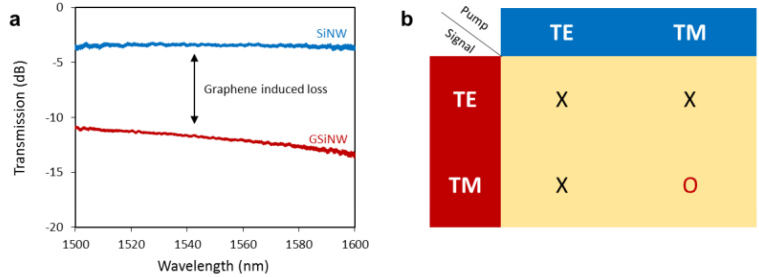

**Figure S5.1 | Transmission and polarization combinations. a,** Continuous-wave signal transmission: silicon nitride waveguide (SiNW) without graphene layers (blue curve), and the GSiNW (red curve). **b,** When pump and signal varies their polarizations, only TM-TM can generate the DFG-based plasmon in graphene.

### S5.2 Pre-saturation of the GSiNW by using CW signal

Figure S5.2a plots the transmission versus the launched power, over four GSiNW samples. Red dots are the measurements with theoretically fitted blue curve and the noise width is denoted by the grey region. Clear saturable absorption of the GSiNW starts from ≈ 100 mW (20 MW/cm$^2$) and the GSiNW is almost fully saturated when the launched power is above 1 W (0.2 GW/cm$^2$). The saturable absorption induced transmission increase is ≈ 63%. Enabled by the saturable absorption [S46], the high power pulsed pump can modulate the low power CW signal.

Figure S5.2b shows the modulated CW signal measured after the balanced photodetector. The launched CW signal and pump powers are 1 mW and 32 mW respectively. The modulated CW is of the same temporal profile and the same repetition rate of the pulsed pump. Hence, the lock-in amplifier cannot filter off the modulation induced signal enhancement. That means, after the lock-in amplifier, the background of the enhanced signal spectrum is not 0. For pristine graphene, the modulation can be three orders of magnitude larger than the DFG based enhancement. When the modulation is too large, it might saturate the detector,

rendering the DFG enhanced peak undetectable. To suppress this modulation, we use high power CW signal to *pre-saturate* the graphene layers. Figure S5.2c shows the lock-in amplified signals, by using the CW laser with 1.2 W, 1.4 W, and 1.6 W powers.

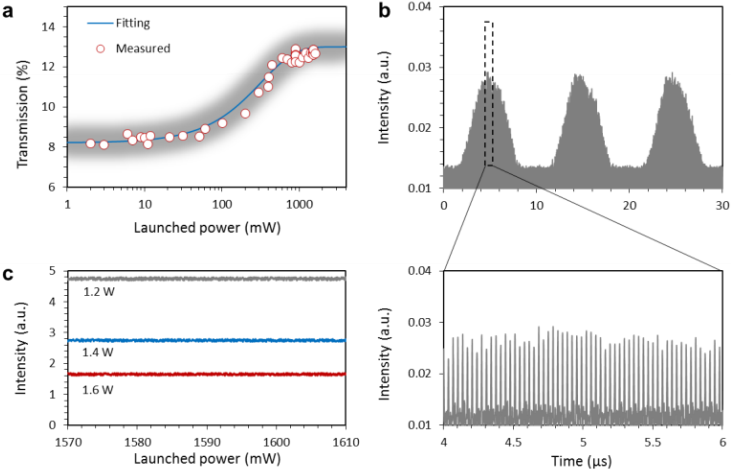

**Figure S5.2 | Saturable absorption induced modulation. a,** Saturable absorption of the GSiNW. **b,** Modulated signal. **c,** Modulation enhanced 100 kHz signal, amplified by the lock-in amplifier, when the CW signal is 1 W (grey), 1.2 W (blue) and 1.4 W (red).

### S5.3 DFG enhanced signal of a GSiNW with 60 nm thick $Al_2O_3$

The $Al_2O_3$ layer thickness not only determines the $V_G$-$I_{SD}$ curve of the graphene-$Al_2O_3$-graphene transistor, but also influences the plasmon coupling. Figure S5.3 shows the DFG enhanced signal at $V_G = 0$ V, when thickness of the $Al_2O_3$ is 60 nm. Compared to the GSiNW with 30 nm thick $Al_2O_3$ (blue curve, peak location 1593.7 nm, $f_{SP}$ = 7.4 THz), the GSiNW with 60 nm thick $Al_2O_3$ (red curve) has a peak location at 1589.9 nm ($f_{SP}$ = 7.1 THz). We regard that there is little plasmon coupling in a 60 nm graphene-$Al_2O_3$-graphene system.

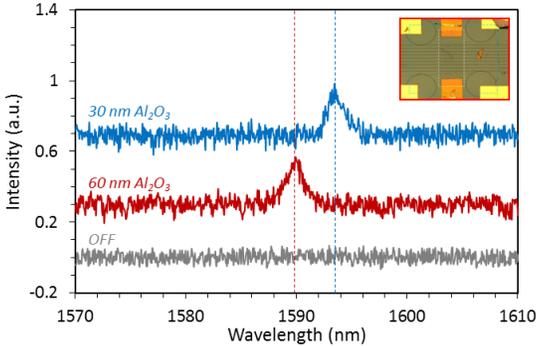

**Figure S5.3 | Enhanced spectra. a,** Grey: pump off; Blue: GSiNW with 30 nm $Al_2O_3$; Red: GSiNW with 60 nm $Al_2O_3$. Inset: Optical micrograph of the GSiNW with 60 nm $Al_2O_3$.

*S5.4 Measurement of the plasmons with pump frequency tuning*

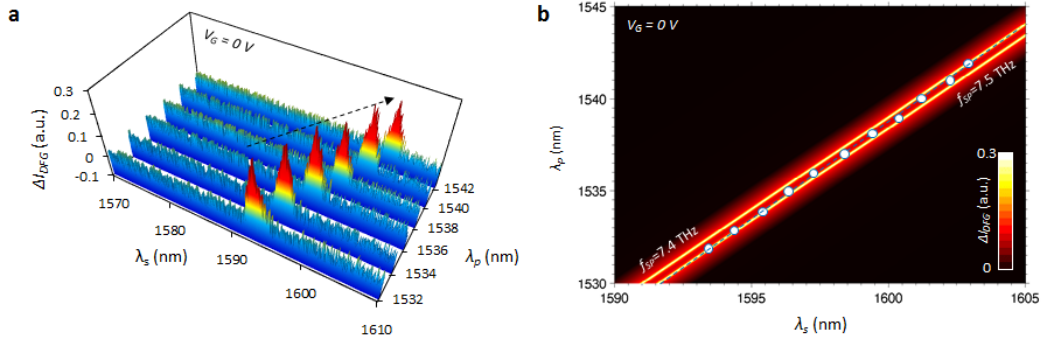

**Figure S5.4 | Tuning the pump frequency. a,** Spectra of the enhanced signal, when $\lambda_p$ is tuned from 1532 nm to 1542 nm. **b,** Measured $\lambda_p$-$\lambda_s$ correlation, with the $f_{SP}$ slightly shifted from 7.5 to 7.4 THz due from dispersion matching. When $\lambda_p$ is 1532 nm, $f_{SP}$ is 7.5 THz; when $\lambda_p$ is 1542 nm, $f_{SP}$ is 7.4 THz.

When $\lambda_p$ is tuned from 1532 nm to 1542 nm (195.8 THz to 194.6 THz), the enhanced signal peak $\lambda_s$ is shifted from 1593.2 nm to 1603 nm (188.3 THz to 187.2 THz) as shown in Figures S5.4a and S5.4b. During this process, $f_{SP}$ decreases from 7.5 THz to 7.4 THz. The trace of the $f_{SP}$ follows the graphene plasmonic dispersion well, as described in the main text.

**Supplementary References:**